\documentclass{elsarticle}

\usepackage{graphicx}
\usepackage{subfigure}
\usepackage{algorithm}
\usepackage{algorithmic}
\usepackage{setspace}
\usepackage{amsmath}
\usepackage{multirow}
\usepackage{amssymb}
\usepackage{color}
\usepackage{lineno,hyperref}
\usepackage{epstopdf}
\usepackage{geometry}
\usepackage[numbers]{natbib}
\journal{}
\begin{document}
\begin{spacing}{1.2}
\begin{frontmatter}

\title{RSBNet: One-Shot Neural Architecture Search for A Backbone Network in Remote Sensing Image Recognition}
\author[mymainaddress]{Cheng Peng}

\author[mymainaddress]{Yangyang Li\corref{mycorrespondingauthor}}
\cortext[mycorrespondingauthor]{Corresponding author}
\ead{yyli@xidian.edu.cn}

\author[mymainaddress]{Ronghua Shang}

\author[mymainaddress]{Licheng Jiao}

\address[mymainaddress]{Key Laboratory of Intelligent Perception and Image Understanding of Ministry of Education, School of Artificial Intelligence, Xidian University, Xi'an, 710071, China}

\begin{abstract}
Recently, a massive number of deep learning based approaches have been successfully applied to various remote sensing image (RSI) recognition tasks. However, most existing advances of deep learning methods in the RSI field heavily rely on the features extracted by the manually designed backbone network, which severely hinders the potential of deep learning models due the complexity of RSI and the limitation of prior knowledge. In this paper, we research a new design paradigm for the backbone architecture in RSI recognition tasks, including scene classification, land-cover classification and object detection. A novel one-shot architecture search framework based on weight-sharing strategy and evolutionary algorithm is proposed, called RSBNet, which consists of three stages: Firstly, a supernet constructed in a layer-wise search space is pretrained on a self-assembled large-scale RSI dataset based on an ensemble single-path training strategy. Next, the pre-trained supernet is equipped with different recognition heads through the switchable recognition module and respectively fine-tuned on the target dataset to obtain task-specific supernet. Finally, we search the optimal backbone architecture for different recognition tasks based on the evolutionary algorithm without any network training. Extensive experiments have been conducted on five benchmark datasets for different recognition tasks, the results show the effectiveness of the proposed search paradigm and demonstrate that the searched backbone is able to flexibly adapt different RSI recognition tasks and achieve impressive performance.
\end{abstract}

\begin{keyword}
Remote sensing image, neural architecture search (NAS), convolutional neural networks (CNNs), evolutionary algorithm
\end{keyword}

\end{frontmatter}
\section{Introduction}\label{section:1}
The earth observation and remote sensing technologies have made remarkable progress in the last few decades, which significantly increased the quantity of remote sensing image (RSI) and enable us to measure and observe the detailed structure of the earth's surface with high resolution aerial images \cite{lateef2019survey}\cite{chen2021adaptive}. Driven by many real-world applications such as urban planning, environment monitoring, and vegetation mapping \cite{wang2019accurate}, the interpretation of RSI has become an active research direction and raised increasing interests. Recently, the emergence of deep learning provides a promising approach to solving the RSI recognition tasks \cite{yi2021probabilistic}\cite{bi2021multi}\cite{tan2020multi}. As a typical and important model in deep learning community, deep convolutional neural networks (CNNs) construct an end-to-end feature extraction framework, which have achieved excellent performance in RSI recognition tasks, and have gradually become the mainstream methods in RSI interpretation field\cite{zhu2020diverse}\cite{li2020parallel}.

\begin{figure}[htbp]
 \centering
 \includegraphics[scale=0.3]{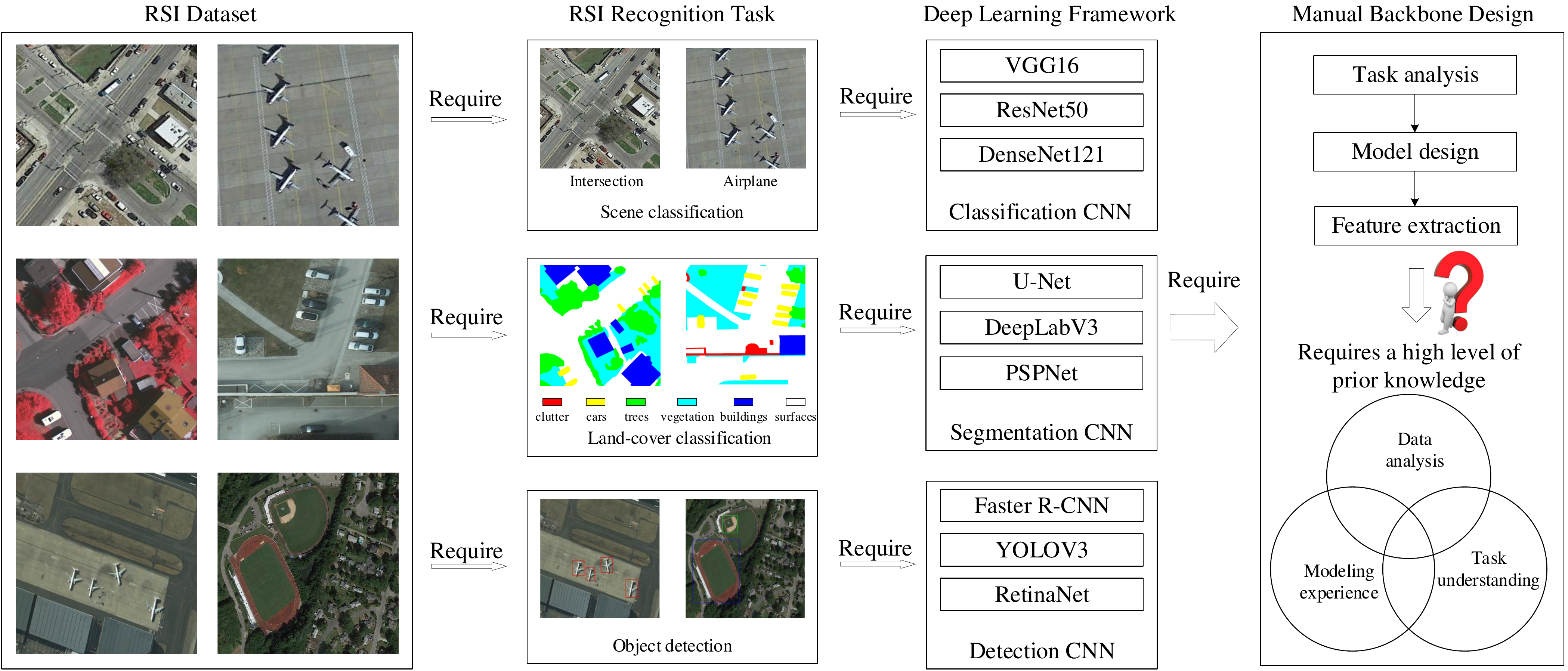}
  \caption{Overview of manually designed CNNs for the RSI recognition task.}
  \label{fig:overview}
\end{figure}

Generally speaking, CNN based methods for the task of RSI analysis contain two basic steps: feature extraction and recognition \cite{peng2020efficient}. The feature extractor in deep neural networks (DNNs) is also called backbone network in many literatures \cite{li2020object}, which plays a significant role in various RSI recognition tasks. As shown in Fig.\ref{fig:overview}, the potential of deep neural networks heavily depends on the representative abilities of the features extracted by backbones, and many breakthroughs in deep learning stem from their immediate modifications. For instance, a large classification accuracy increase could be obtained by simply replacing a VGG-16 \cite{simonyan2014very} backbone network with stronger networks, \emph{e.g.}, ResNet-50 or ResNet-101 \cite{he2016deep}. There has been a long time that most promising backbone such as VGG \cite{simonyan2014very}, ResNet \cite{he2016deep}, MoblieNet \cite{howard2017mobilenets} and SENet \cite{hu2018squeeze} are manually designed by the researchers with a high level of prior knowledge. As a black-box optimization task in a discrete search space, the design of neural architectures usually requires extensive computation and time cost, and may take many trials before achieving satisfied results. In addition, the handcrafted neural architectures are often task-dependent, which means the designed neural networks are difficult to adapt different distributions and different recognition tasks. Many excellent DNNs architecture are originally designed for natural image classification \cite{tan2019efficientnet}. However, the significant differences between RSI and natural images should be noticed. RSIs generally focus on the roof information of the ground objects which have different orientations, scales, and spatial distributions, whereas the natural images usually contain the profile information of the objects \cite{li2020object}. Currently, many researches for RSI recognition directly employ the networks designed for natural image classification as backbones, which might be sub-optimal due the significant domain differences. Therefore, it is urgent to design special backbone architecture for different RSI recognition tasks.

More recently, neural architecture search (NAS) methods have emerged in the community of automated machine learning (AutoML) \cite{he2021automl}, which aims to reduce the manual intervention of such architecture engineering, and have achieved promising results in various tasks such as image classification and segmentation. The goal of NAS methods is to automatically find the optimal architecture through choosing and combining various neuronal operations from a predefined search space. Nested optimization is usually adopted in early NAS approaches, where a massive number of candidate architectures are sampled from the search space using reinforcement learning (RL)\cite{baker2016designing} or evolutionary algorithm (EA)\cite{real2019regularized}, and the network weights are then trained from scratch to validate the performance, leading to unaffordable computation cost (e.g., thousands of GPU days). Consequently, a lot of methods have been presented to reduce the computation cost and accelerate the search process. Gradient-based methods (\emph{e.g.}, DARTS \cite{liu2018darts}) are developed in recent works, which construct an over-parameterized network that contains all candidate connections and introduce the attention mechanism on each candidate operation. Thus the discrete search space is transformed to be continuous and the architecture parameters can be optimized by gradient descent methods, which can significantly reduce the search cost while keeping competitive performance. Other NAS methods (e.g., ENAS \cite{pham2018efficient}) propose to introduce parameter sharing strategy to reduce the computation cost of architecture evaluation \cite{wu2019fbnet}. As a recently developed paradigm based on weight sharing strategy, one-shot approaches have raised increasing interests due to the efficiency and flexibility \cite{guo2020single}. A supernet that contains all architectures is defined in one-shot search framework, and then the supernet is trained only once and share the network weights with its subnet. Therefore, the evaluation cost of subnet is more efficient and affordable on large datasets.

Despite the remarkable performance of NAS in many image understanding tasks and the related research works are complicated and rich, most existing NAS methods are studied based on the task of natural image classification, few researches have investigated the performance of NAS in the RSI recognition tasks including scene classification, land-cover classification and object detection. In addition, most public RSI datasets contain insufficient labeled data and have some significant disadvantages, such as the few scene and object categories, and the low image diversity. Currently, various pretrained DNNs (e.g., VGG \cite{simonyan2014very}, GoogLeNet \cite{szegedy2015going}, ResNet \cite{he2016deep}) based on ImageNet \cite{krizhevsky2012imagenet} have been widely adopted in many RSI recognition tasks to address the issue of training data shortage, where the weights of feature extractor in different recognition tasks are initialized as the pretrained weights. However, a recent research \cite{he2019rethinking} has demonstrated that pretraining-based paradigm might be sub-optimal to solve the problem of insufficient training data. Moreover, the training data from aerial sensors may exhibit complex geometrical structures compared with natural images, which affects the performance of pretrained model for RSI recognition.

For image classification, the networks discovered by NAS methods reach or even surpass the performance of the hand-crafted networks. However, NAS for backbone networks in RSI recognition tasks is an open problem deserved to further research. In summary, there are three difficulties should be addressed: (1) optimization difficulty: A complete DNN architecture for recognition usually consists of backbone and classifier. The architecture search problem is deeply coupled during the optimization of other parameters since the NAS framework requires accuracies of candidate architecture on target task. (2) transfer difficulty: NAS system is dataset-specific and task-specific especially for remote sensing images, the searched backbone based on a single small scale dataset is difficult to adapt to different domain information and cannot be transferred to other recognition tasks. (3) optimization inefficiency: Considering the insufficiency of training data, each candidate architecture during search is generally pretrained on ImageNet and then fine-tuned on the target dataset to obtain precious evaluation, which is time-consuming and inefficient. In this paper, we focus on the performance bottlenecks of the manually designed backbone for the task of RSI recognition. We research a new design paradigm for the backbone architecture in the target recognition task. The main contributions of this paper are described as follows.
\begin{enumerate}[(1)]
  \item
  We propose a novel one-shot backbone architecture search framework based on the weight-sharing strategy, where the training process and the search process are decoupled. Our framework consists of three stages: one-shot supernet pre-training, supernet fine-tuning and candidate backbone search based on evolutionary algorithm.
  \item
  We introduce an effective one-shot supernet training strategy, called ensemble single-path training. In each training step, a mini batch of single-path subnetworks are random sampled, the gradients of which are then calculated and accumulated to update the weights of supernet. This strategy accelerates and stabilizes the training process, and guarantees a much high correlation between the training and search stage.
  \item
  We design a switchable recognition module, which can flexibly adapt to different RSI recognition tasks. In the fine-tuning stage, the backbone network is separated from the pretrained supernet and equipped with different recognition head to form different recognition models, which aims to improve the performance of the supernet for the target task.
  \item
  We prove the effectiveness of RSBNet on three RSI recognition tasks, including scene classification, land-cover classification and object detection, based on five public benchmark datasets. Extensive experimental results demonstrate that the searched backbones indicate better generalization ability compared with the manually designed backbones.
\end{enumerate}

The rest of this paper are divided into five sections. Section \ref{section:2} presents a brief review to the related works including deep learning based RSI recognition methods and NAS methods. In \ref{section:3} , we give a detailed description of our proposed RSBNet. Sections \ref{section:4}  and \ref{section:5}  deal with the experimental settings and the comparison results respectively. Our conclusions are discussed in \ref{section:6} .

\section{Related Work}\label{section:2}
Over the past years, numerous efforts have been devoted to the development of various methods for RSI recognition including scene classification, land-cover classification and object detection. Due to the excellent feature extraction ability, deep learning based approaches have become the most popular methods in recent years. However, the design of the deep neural network (DNN) architecture has proven to be significant and challenging. The emergence of NAS methods has raised up a new idea for automatically design the architectures of DNN models. In this section, we present a brief description of related works, including various deep learning based RSI recognition approaches and existing NAS frameworks.

\subsection{Deep Learning Methods in RSI Recognition}\label{section:2.1}
According to the spatial resolution, remote sensing image recognition task can be divided into three parallel classification branches at different levels: pixel-level (i.e., land-cover classification), object-level (object detection), and scene-level (scene classification). Although these sub-tasks have different goals, they share some high-level semantic information. More recently, various studies have shown the impressive performance of deep learning models in RSI recognition. Below we introduce many representative deep learning based methods in RSI recognition from three aspects.

In terms of scene classification task, it aims to capture the global feature of the RSIs. Early RSI scene classification approaches based on deep learning prefer to fully train a new DNN model with random initial weights. For example, Zhang \emph{et al}.\cite{zhang2015scene} developed a gradient boosting random convolutional network (GBRCN) to extract discriminative high-level semantic information from small-scale data. Liu \emph{et al}.\cite{liu2017scene} constructed a triplet network based on weakly supervised training. However, the performance of DNN models heavily depends on the mount of training data, which make it impracticable of training a robust DNN model on limited RSI data. To solve the overfitting problem, some works directly take the activations from the fully-connected layers of pretrained CNNs as the image representations and achieve promising performance \cite{cheng2017remote}. The others employ the DNN models as local feature extractor and incorporate unsupervised methods for feature coding to obtain scene representation. Some methods introduce the idea of feature fusion to generate more discriminative feature representation. Zhao \emph{et al}.\cite{zhao2016spectral} proposed to combine the multispectral information and the structural feature of scene images to boost the performance. Recently, the performance improvement through deep learning based methods for RSI scene classification has gradually reached a bottleneck due to the heavy dependence on the pretrained weights and limited public data scale.

Different from scene classification tasks, land-cover classification is a fine-grained recognition task, which focus on both global contextual semantic information and local detailed features. The emergence of fully convolutional network (FCN)\cite{long2015fully} without any fully connected layers promotes the development of encoder-decoder CNN architectures for end-to-end dense predictions, and provides semantic segmentation tasks a standard paradigm for all the subsequent advanced methods, such as U-Net \cite{ronneberger2015u}, SegNet\cite{badrinarayanan2017segnet} and DeepLab series \cite{chen2018encoder}. Similarly, FCN-based methods have also been widely applied to RSI land-cover classification. For instance, Kuo \emph{et al}.\cite{kuo2018deep} proposed an aggregation feature decoder incorporate with DeepLabV3 model to progressively extract and fuse different-level features from the encoder. Peng \emph{et al}. \cite{peng2019densely} constructed a multiscale and multimodal FCN model to combine both the spatial semantic feature and the information of digital surface models, and investigate early and late fusion strategies through densely connected layers. Li \emph{et al}.\cite{li2019adaptive} developed an adaptive fusion module to combine the feature maps from multiple levels of the backbone network. In order to address the specific difficulties appeared in semantic segmentation task such as reduced spatial resolution caused by pooling layers, blurred edges of different objects and multi scale objects, most existing methods for land-cover classification focus on improving the feature representation abilities of the backbone to adapt specific spatial patterns of remote sensing images.

For a long time, object detection plays an essential role in the field of RSI interpretation. Generally, most existing deep learning methods designed for object detection generally consist of two branches on the basis of whether generating region proposals. One is a two-stage detection framework such as Faster R-CNN \cite{ren2016faster}, Mask R-CNN \cite{he2017mask} and Cascade R-CNN \cite{cai2018cascade}, which first generate a series of candidate region proposals (called anchors) and the classification and regression for the anchors are then performed to predict the bounding box of object. The second branch solve the object detection task in one stage through incorporating the regression and classification process (\emph{e.g.},YOLO series\cite{redmon2018yolov3}, RetinaNet\cite{lin2017focal}.). For instance, in the model Single-Shot multibox Detector (SSD) \cite{liu2016ssd}, the bounding boxes together with probability scores are directly predicted based on a set of default boxes with dense distribution. Inspired by the remarkable performance of deep learning based methods for object detection in natural images, various researchers propose to adapt the state-of-the-art object detectors designed for natural images to remote sensing images. Zhu \emph{et al.}\cite{zhu2020diverse}proposed a multi-branch conditional generative adversarial network to augment data for RSI object detection. Li \emph{et al.}\cite{li2020parallel} introduced a novel parallel down-up fusion network for salient object detection in RSIs, which aims to distinguish diversely scaled salient objects and suppress the cluttered backgrounds. Numerous successful results have demonstrated that CNN architecture serve as backbone network embedded in various object detection frameworks. However, most outstanding feature extractors are originally designed for natural images, the challenges caused by the domain discrepancies should to be addressed.

\subsection{Neural Architecture Search Methods}\label{section:2.2}
Over the past years, many neural architecture search (NAS) methods have been proposed to reduce the manual intervention of the architecture engineering , and have achieved promising success in various image recognition tasks such as image classification and object detection. Early NAS methods employ various sampling strategies based on reinforcement learning (RL) \cite{baker2016designing} or evolutionary algorithm (EA) \cite{real2019regularized} to sample a large number of candidate architectures from the search space and their network weights are then trained from scratch to validate the performance, which consumes large computational overhead. Recent differentiable architecture search (DARTS) method has been proposed as an alternative \cite{liu2018darts}, which designed a continuous search space and constructs an over-parameterized network through introducing the attention mechanism to each connection in the architecture. The supernet weights and the architecture parameters are jointly optimized based on gradient decent, resulting in orders of magnitude fewer search cost compared with RL-based and EA-based approaches. Despite the excellent performance of differentiable architecture search method, there are two issues should be noticed. First, the weights in the supernet are deeply coupled with the architecture parameters, which makes the gradient evaluation prohibitive due to the expensive inner optimization. Second, optimization bias of architecture distribution is inevitably introduced during the search process due the greedy nature of the gradient based methods, leading to local trap or overfitting phenomenon for the architecture search. As a new NAS paradigm, one-shot methods introduce the strategy of weight sharing \cite{bender2018understanding}, which trains a supernet that subsumes all architectures and derives child architecture performance from the supernet without any training. Compared with DARTS framework, there is no search space relaxation and architecture distribution parameterization in one-shot architecture method. Thus one-shot framework usually contains two stage: supernet training and subnet search, which is more flexible to adapt different recognition task. To the best of our knowledge, most of the existing NAS methods for image recognition are designed for natural images, few literatures have investigated the performance of NAS for the task of RSI recognition.

\section{Methodology}\label{section:3}
\begin{figure}[htbp]
 \centering
 \includegraphics[scale=0.24]{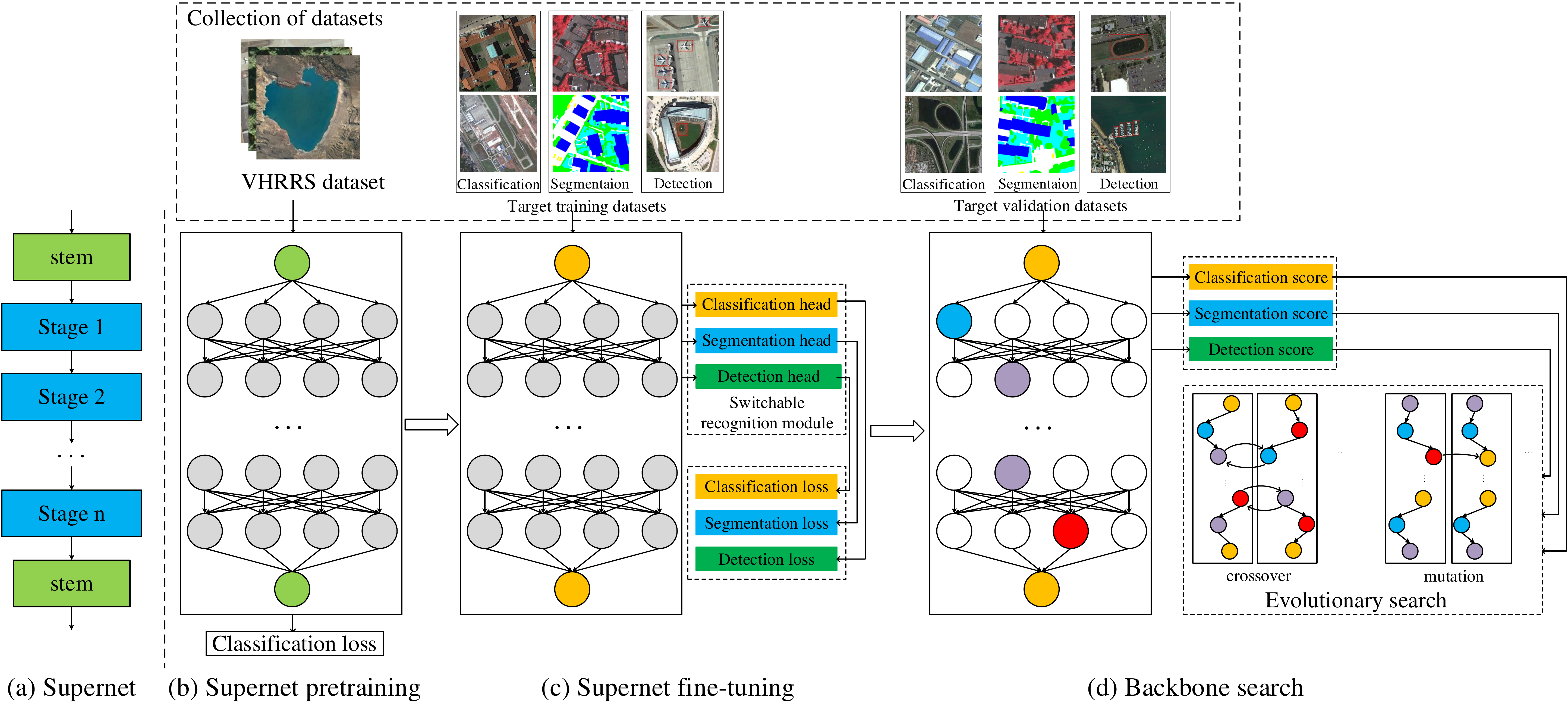}
  \caption{Overall framework of the proposed RSBNet, which contains three stage:(1)one-shot supernet pretraining; (2)supernet fine-tuning for target task; (3)backbone search with evolutionary algorithm.}
  \label{fig:framework}
\end{figure}

In this work, we present a new design paradigm for the backbone network architecture in DNN models for remote sensing image recognition. The proposed RSBNet contains three basic stage:
\begin{enumerate}[(1)]
\item
  One-shot supernet pretraining: We first construct a supernet in a predefined search space, which contains a wide variety of candidate architectures. Then we perform gradient-based training for the one-shot model over a self-assembled large-scale RSI dataset (see section \ref{section:4.2} ), which aims to improve the classification accuracy of all sub-networks that are derived from the supernet.
\item
  Task-specific supernet fine-tuning: We design a switchable recognition module that can flexibly adapt to different recognition tasks. Specifically, the backbone network is separated from the pretrained supernet of the first stage and equipped with corresponding classifier to form different recognition models. Then we respectively fine-tune these different supernets over the target dataset to obtain task-specific supernet.
\item
  Backbone search: We search for the optimal backbone architecture through selecting the paths in the supernet under the direction of the evolutionary controller. Noted that the weights of subnetwork are inherited from the supernet without any training. Besides, we can obtain different architectures that stably meet hard computational constraints, \emph{e.g.}, FLOPs and parameter size.
\end{enumerate}
The overall pipeline of our method for RSI recognition is illustrated at Fig.\ref{fig:framework}. Algorithm \ref{alg:framework} shows the detailed steps. Below we provide a detailed description of our methods.

\begin{algorithm}[htb]
\caption{Framework of the proposed RSBNet}
\label{alg:framework}
\begin{algorithmic}[1]
\REQUIRE supernet $N({\cal A},W)$, pretraining dataset ${D_{pretraining}}$, fine-tuning dataset ${D_{fine-tuning}}$, searching dataset ${D_{searching}}$, pretraining epochs ${S_{pretraining}}$, fine-tuning epochs ${S_{fine-tuning}}$, maximum number of evolutionary generations $G$, the number of subnetworks $B$ sampled in each step;
\STATE Stage 1: Supernet pretraing

\FOR{$i = 1, ...,{S_{pretraining}}$ }
\FOR{data $X$, label $Y$ in ${D_{pretraining}}$ loader}
\STATE Randomly sample $B$ single-path subnetworks from the supernet $N({\cal A},W)$;
\STATE Execute $B$ sampled subnetworks;
\STATE Calculate gradients for each sampled subnetwork $N(a,w)$;
\STATE Update the supernet weights $W$ according to \eqref{eq4} by accumulated gradients;
\ENDFOR
\ENDFOR
\STATE Stage 2: Supernet fine-tuning
\STATE Construct the supernet for the target task through the switchable recognition module;
\FOR{$j = 1, ...,{S_{fine-tuning}}$ }
\STATE Fine-tuning the supernet on the fine-tuning dataset ${D_{fine-tuning}}$ according to Algorithm \ref{alg:finetune};
\ENDFOR
\STATE Stage 3:Backbone search
\FOR{$k = 1, ...,G$ }
\STATE Evolutionary search the backbone for the target task based on the searching dataset ${D_{searching}}$;
\ENDFOR
\ENSURE The optimal backbone for the target task.
\end{algorithmic}
\end{algorithm}

\subsection{One-shot Supernet Search Space}\label{section:3.1}
For search efficiency, many previous NAS methods focus on cell-based search space, which generally contain two types of building blocks. (\emph{i.e.}, normal cell and reduction cell). Once the cell structures are learned, they are repeatedly stacked to construct the final architecture. However, simply forcing all the cells share the same architecture hinders the potential of deep neural networks since the same cell architecture at different layers have different representation ability.

\begin{figure}[htb]
\centering
\includegraphics[scale=0.8]{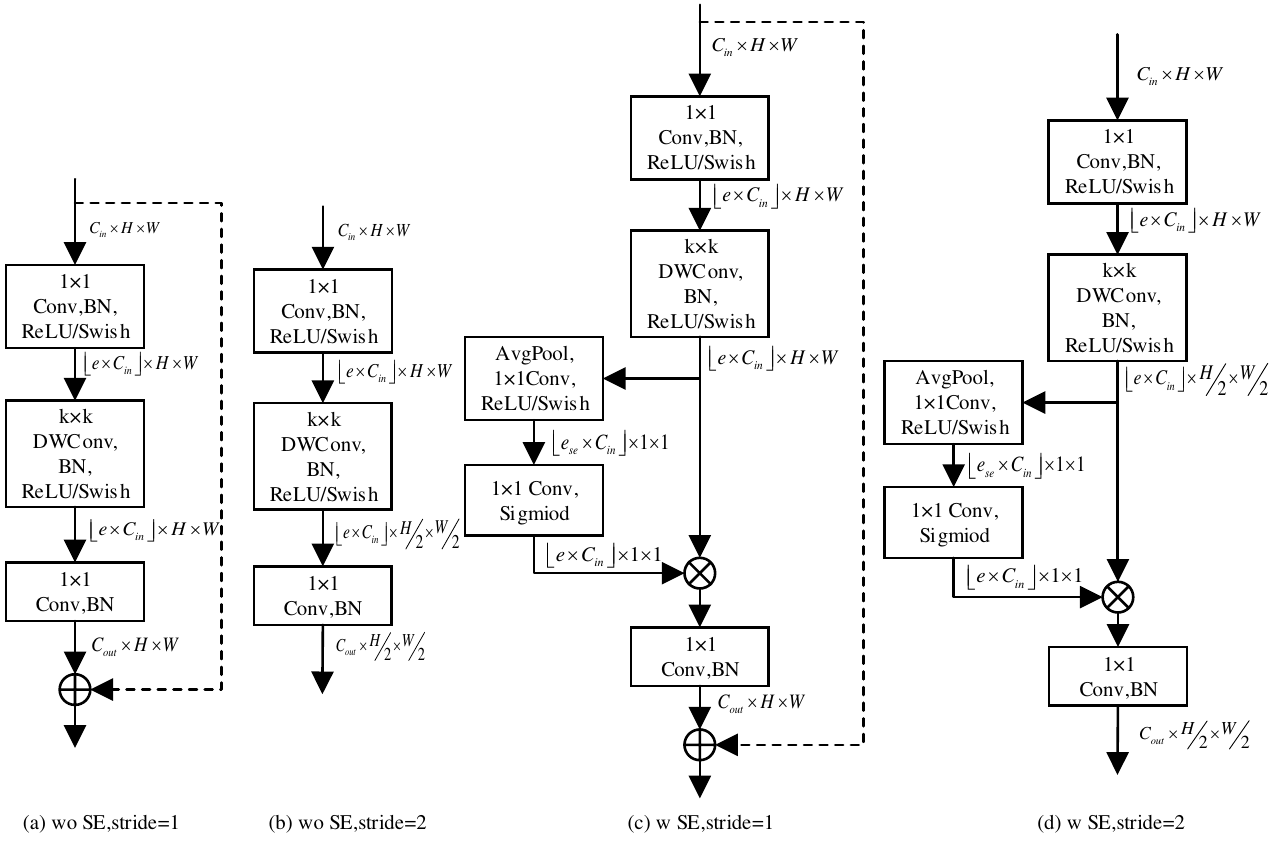}
\caption{Illustrations of candidate block architectures.}
\label{fig:block}
\end{figure}

\begin{table}[htb]
\centering
\footnotesize
\renewcommand\arraystretch{1.2}
\renewcommand\tabcolsep{3.5pt}
\caption{The overall architecture of the supernet}
\begin{tabular}{c|c|c|c|c|c|c}
  \hline

  Stage&Input&Operation& ${C_{out}}$&Act. & Expansion& n \\
  \hline
  1 & ${224^2} \times 3$ &$3 \times 3$ conv& 32 & ReLU & - & 1 \\
  2 & ${112^2} \times 32$ &MBInvRes & 16 & ReLU & 1 & 1 \\
  3 & ${112^2} \times 16$&OPS & 24 & ReLU & [2, 6]& 4 \\
  4 &${56^2} \times 24$ &OPS& 40 & Swish & [2, 6]& 4\\
  5 &${28^2} \times 40$&OPS & 80 & Swish & [2, 6]& 4\\
  6& ${14^2} \times 80$&OPS & 112 & Swish & [2, 6] & 4\\
  7 & ${14^2} \times 112$&OPS & 192 & Swish & [2, 6]& 4\\
  8 & ${7^2} \times 192$&OPS & 320 & Swish & [2, 6]& 1\\
  9 & ${7^2} \times 320$& $1 \times 1$ conv & 1280 & Swish & -& 1\\
  10 & ${7^2} \times 1280$&Avgpool & 1280 & -& -& 1\\
  11 &1280&Fc & 1000 & - & -& 1\\
  \hline
\end{tabular}
\label{tab:archi}
\end{table}
Our one-shot network is constructed based on a layer-wise search space, which consist of a fixed number of layers, and the building block in each layer is searchable. Table \ref{tab:archi} shows the detailed configuration of overall architecture, which defines the number of layers and the resolution of feature maps in each layer.  The first two layers of the network contains fixed blocks. The intermediate layers are divided into multiple stages, each stage contains predefined number of layers and their block type is chosen from the block search space. ``OPS'' in third column denotes the building blocks to be searched, ``$C_{out}$'' denotes the number of output channels, ``Act'' represents the activation function, ``n'' is the number of layers in a stage, ``Expansion'' defines the width of a layer and [a, b] is a discrete interval. The basic candidate blocks is the basic block in MobileNetV2 (``MBInvRes'') \cite{sandler2018mobilenetv2}. To enhance the representative ability of the network, we introduce the channel attention module (Squeeze-and-Excitation module) \cite{hu2018squeeze} into the candidate blocks. Notably, the supernet is trained with full width in the pretraining stage, which in the fine-tuning stage, we optimize the supernet with elastic width, that is to say, we allow the number of feature map channels (i.e., the expansion ratio of the depthwise separable convolution in MBInvRes block) to be searchable to support more complex search spaces. In our search space, each candidate block has a kernel size $k = 3$ or $k = 5$, and a discrete expansion ratio $e \in [2,6]$ (with stride 0.5) for the depthwise convolution. The expansion ratio in SE module is set to 1. The detailed architecture of candidate blocks are illustrated in Fig.\ref{fig:block}.

\subsection{One-shot Supernet Pretraining}\label{section:3.2}
In this stage, we focus on training a robust one-shot supernet and improving the classification performance of all subnetworks that are derived by selecting different connections from the supernet. Without loss of generality, our one-shot NAS framework first trains a supernet once and allows all candidate architectures to share the weights without any training. We denote the architecture search space as ${\cal A}$ and a supernet with network weights $W$ is represented as $N({\cal A},W)$. If the supernet architecture is constructed, the network weights $W$ can be optimized by standard gradient decent, which fits for all the subnetworks $N(a, w)$ to achieve excellent classification performance.

\begin{equation}
{W^{\rm{*}}} = \mathop {\arg \min }\limits_W {{\cal L}_{train}}(N({\cal A},W))
\end{equation}

Once the training loss are converged, we could consequently sample the subnetworks for instant evaluation without any training. However, training the one-shot supernet is a non-trivial and computationally prohibitive task due to the joint optimization of the weights, which requires obtain the exact gradient in each optimization step to maintain the classification accuracy of a massive number of subnetworks. For example, previous studies \cite{bender2018understanding} found that the training stage is unstable and the training loss may explode if the one-shot supernet is not properly initialized. Moreover, removing connections (even unimportant ones) after the training process from the supernet can cause the severe degradation of the subnetwork performance, which shows a low correlation between the supernet accuracy and the stand-alone model performance.

To alleviate the coupling phenomenon during training stage and make the training process more efficient, we propose a simple and efficient ensemble single-path training strategy, where the supernet weights are optimized in a way that only the weights of random sampled single-path subnetworks are simultaneously updated in each step. Let us restart to derive the fundamental principle of supernet training. We denote a subnetwork ${N_i}$ ($i = 1,...,N$, $N$ is the number of candidate subnetworks) equipped with weights ${W_i}$ as $N(C_i, W_i)$, where $C_i$ represents a set of binary connection parameters (\emph{i.e.}, $\{0, 1\}$) and $W_i$ is the network weights of $N_i$. The 0-element in $C_i$ means the corresponding block and weights are not contained in the subnetwork, and 1-element indicates the corresponding block is contained in the subnetwork. From this point of view, we have ${W_i} = W \odot C_i$ ($W$ is the collection of all the weights in the supernet, $ \odot $ is a mask operation which maintain the weights of the supernet only for the corresponding blocks equal to 1).

At each training step, a mini-batch of training data $X$ are simultaneously fed into the $M$ random sampled subnetworks, the predictions of each network is $N_i(X)$. The training loss can be calculated as ${{\cal L}_i} = F({N_i}(X),Y)$, where $F$ is the loss metric and $Y$ is the truth label. Thus the gradient of $W_i$ in $N_i$ can be expressed as
\begin{equation}
d{W_i} = \frac{{\partial {{\cal L}_i}}}{{\partial {W_i}}} = \frac{{\partial {{\cal L}_i}}}{{\partial W}} \odot {C_i}
\end{equation}

In a naive way, the number of sampled subnetworks $M$ should be the size of search space, and the model weights $W$ should fit all the subnetworks. Consequently, the gradients of the supernet can be calculated through accumulating all subnetworks gradients, and updated based on standard gradient decent
\begin{equation}
dW = \frac{1}{M}\sum\limits_{i = 1}^M {d{W_i}}  = \frac{1}{M}\sum\limits_{i = 1}^M {\frac{{\partial {{\cal L}_i}}}{{\partial W}} \odot C}
\label{eq3}
\end{equation}

The weights of a block are only optimized when the subnetwork contains this block during forwarding. Unfortunately, this training principle is time-consuming and infeasible for large-scale search space since we collect a large number of candidate architectures with shared weights in the supernet. We propose a simple but efficient training principle called ensemble single-path training strategy, which is inspired by the idea of stochastic gradient descent (SGD). We only employ a mini-batch of subnetworks for updating the supernet weights $W$. Specifically, we employ uniform sampling without replacement and sample $B$ candidate subnetworks at each step, where $B \ll M$ and the sampled architecture only contains single candidate block in each layer. The parameter updating principle of \eqref{eq3} is transformed as
\begin{equation}
dW \approx \frac{1}{B}\sum\limits_{j = 1}^B {\frac{{\partial {{\cal L}_i}}}{{\partial {W_j}}}}
\label{eq4}
\end{equation}

\begin{figure}[htbp]
\centering
\includegraphics[scale=0.35]{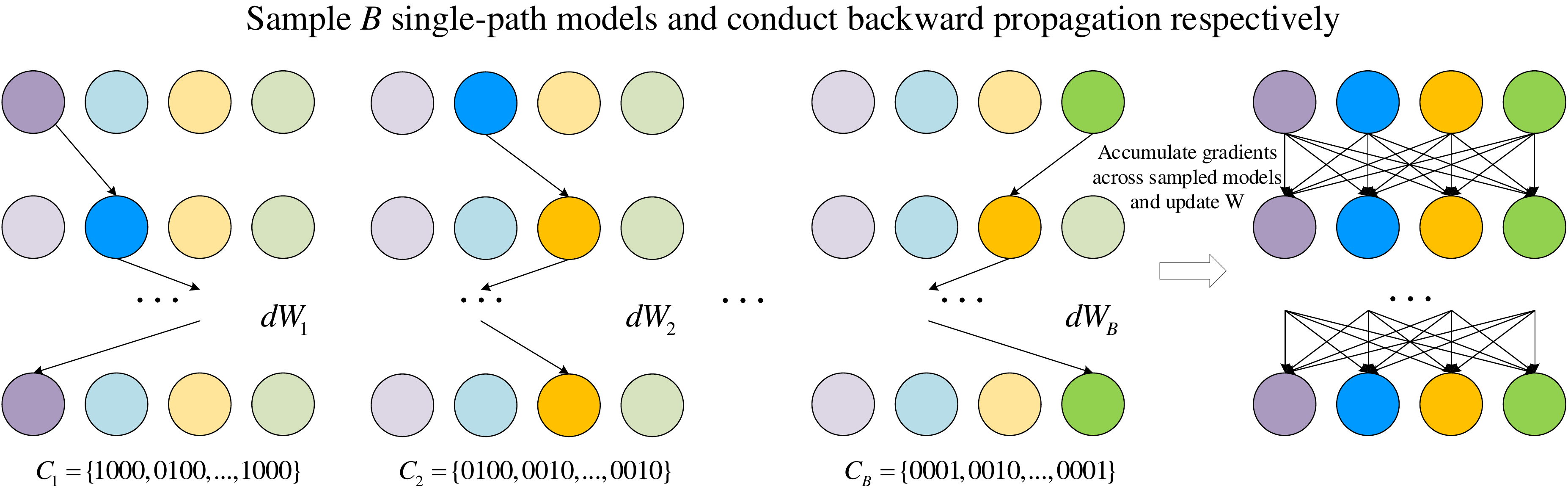}
\caption{Our ensemble single-path supernet training strategy in each step.}
\label{fig:train}
\end{figure}

Consequently, the gradients over a mini-batch of sampled subnetworks are accumulated as an unbiased approximation of the averaged gradients of all the $B$ different candidate architectures. Notably, our method is different from the work in \cite{guo2020single}, which uniformly sample only one single-path network (only a candidate block in each layer) in each step, and perform back-propagation (BP) operations immediately for the sampled subnetworks after the forwarding. In our method, $B$ single-path architectures are simultaneously performed back-propagation. As shown in Fig.\ref{fig:train}, the gradients are then accumulated across the sampled architectures but the weights of supernet are updated when all $B$ BPs are finished. Therefore, the computation cost for each update of supernet could be significantly reduced compared with naive training.

Considering the significant discrepancies between RSIs and natural images, simply training the one-shot supernet over the large scale natural image dataset ImageNet is infeasible and inefficient. We propose a mergence strategy to build a new large-scale remote sensing image dataset called VHRRS dataset (see section \ref{section:4.2} for details), which contains a diversity of remote sensing scene images with different domain information, making it feasible to train a robust one-shot supernet with strong generalization ability for different RSI recognition tasks. In the stage of one-shot supernet training, the training data is selected from VHRRS dataset.
\subsection{One-shot Supernet Fine-tuning}\label{section:3.3}
In this stage, we design a switchable recognition module, which allows the pretrained one-shot supernet to automatically adapt different RSI recognition tasks including scene classification, land-cover classification and object detection. Considering that the one-shot supernet is trained on a self-assembled large-scale dataset (see section \ref{section:4.2}  for details), directly using the network designed for image classification as the backbone of different recognition tasks might be sub-optimal, because the model trained for image classification focus on global semantic feature, while the task of land-cover classification and object detection not only focus on global contextual semantic information but also on local detailed features.

For different recognition tasks on the target dataset, we can formulate the supernet fine-tuning process as follow
\begin{equation}
\begin{split}
{W^{\rm{*}}}{\rm{ = }}\mathop {\arg \min }\limits_{W \leftarrow {W_p}} {\cal L}_{{\rm{train}}}^{{\rm{  target}}}(N({\cal A},W))\\
{\rm{s}}{\kern 1pt} {\rm{.t}}{\rm{.}}{\kern 1pt} {\kern 1pt} {\kern 1pt} {\kern 1pt} {\kern 1pt} {\kern 1pt} {\kern 1pt} {\kern 1pt} {\kern 1pt} W_p^* = \mathop {\arg \min }\limits_{{W_p}} {\cal L}_{{\rm{train}}}^{{\rm{ cls}}}(N({\cal A},W))
\end{split}
\end{equation}
where ${W \leftarrow {W_p}}$ is to optimize the supernet weights $W$ with ${\kern 1pt} W_p^*$ as initialization. The pre-trained weights ${\kern 1pt} W_p^*$ are necessary for ${W^*}$ since the excellent representative semantic features are generated based on the supernet trained on a large-scale RSI dataset, which facilitate the process of fine-tuning. The fine-tuning of supernet also adopts ensemble single-path training strategy but equipped with different recognition head, metrics and datasets. It is worth mentioning that the supernet is independently fine-tuned on different tasks. Specifically, the switchable recognition module consists of three branches:

\begin{itemize}
  \item
  \textbf{Scene classification branch}: Since the one-shot supernet are pre-trained for scene classification task, the supernet can be directly employed as the feature extractor and finetuned on the target dataset for scene classification task.
  \item
 \textbf{Land-cover classification branch}: For the land-cover classification tasks, we adapt the pretrained one-shot supernet to the semantic segmentation by applying dilated convolution to extract dense features and increase the receptive field. Inspired by \cite{chen2018encoder}, we introduce the atrous spatial pyramid pooling (ASPP) module to replace the classifier layers in the pretrained supernet, which aims to capture multi-scale context and boost the segmentation performance for complex ground objects.
  \item
  \textbf{Object detection branch}: For the task of object detection, the one-shot pretrained supernet can be embedded in different object detection framework, including one-stage framework (\emph{e.g.,} RetinaNet) and two-stage framework (\emph{e.g., }Faster R-CNN),through replacing the classifier layers with corresponding detection head.
\end{itemize}

\begin{algorithm}[htb]
\caption{Fine-tuning for target task}
\label{alg:finetune}
\begin{algorithmic}[1]
\REQUIRE Pretrined supernet $N({\cal A},W)$, target dataset ${D_{target}}$, fine-tuning epochs ${S_{fine-tuning}}$, predefined width list, for example,[2,6] with stride of 0.5, the number of subnetworks $B$ sampled in each step;
\FOR{$i = 1, ...,{S_{fine-tuning}}$ }
\FOR{data $X$, label $Y$ in ${D_{target}}$ loader}
\STATE Randomly sample $B$ single-path subnetworks from the pretrained supernet $N({\cal A},W)$;
\STATE Execute $B$ subnetworks with the smallest width and the largest width respectively;
\STATE Accumulate gradients, \emph{loss.bakcward()};
\FOR{\emph{subnetwork} in\emph{ sampled subnetworks}}
\STATE Randomly select a \emph{width} from the width list;
\STATE Execute the \emph{subnetwork} at \emph{width};
\STATE Accumulate gradients, \emph{loss.bakcward()};
\ENDFOR
\ENDFOR
\STATE Update weights $W$;
\ENDFOR
\ENSURE The fine-tuned supernet for the target task.
\end{algorithmic}
\end{algorithm}

Besides the topology of the supernet, we also allow different configurations of the layer width (\emph{i.e.}, the expansion ratio of the depthwise separable convolution in candidate block) during the supernet fine-tuning stage to support more complex search space, and to satisfy different computational constraints. However, searching the width of convolutional layers is non-trivial since the number of output channels in the current layer is correlated with the number of input channels in the next layer. Inspired by \cite{yu2018slimmable}, we train a slimmable supernet based on the ``sandwich'' training rules. In each step, the width of each layer in the sampled subnetwork are selected from a predefined width range, for example $e \in [2,6]$ (with stride 0.5). We also follow the approaches in \cite{yu2019universally} to address the issue of batch normalization (BN) statistics inconsistency. Similarly, the parameters of all subnetworks with different widths are shared and the active channels in different layers can be correspondingly adjusted.

\subsection{Candidate Backbone Network Search}\label{section:3.4}
In third stage, backbone architecture search is respectively performed on the finetuned supernet from the three different branches of the switchable recognition module. The goal for backbone architecture search is to find the optimal backbone ${a^*}$ that maximizes the validation accuracy of target dataset and satisfies the computational constraints
\begin{equation}
\mathop {\max }\limits_{a \in {\cal A}} ACC_{{\rm{val}}}^{{\rm{target}}}(N(a,{W_{\cal A}}(a))),{\kern 1pt} {\kern 1pt} {\kern 1pt} {\kern 1pt} {\kern 1pt} {\kern 1pt} {\kern 1pt} {\rm{s}}{\rm{.t}}{\rm{.}}{\kern 1pt} {\kern 1pt} {\kern 1pt} {\kern 1pt} {\kern 1pt} {\kern 1pt} {\cal T}(a) < \Omega
\end{equation}
where the weights associated with the sampled subnetwork are inherited from the finetuned supernet. Evaluation of $ACC_{{\rm{val}}}^{{\rm{target}}}$ only requires inference without any retraining since the architecture weights are properly initialized. Therefore, the search is very efficient and flexible, and can be solved by any search algorithm. ${\cal T}( \cdot )$ is a measurement of the resource consumption for the searched architecture, such as parameter size and FLOPs ,and $\Omega $ is the target constraint.

Due to the excellent global search capability and flexibility in dealing with different computational constraints, the evolutionary algorithm is used to search the optimal backbone network in our experiments. We construct and maintain a candidate architecture population $C = \{ {C_1},...,{C_K}\} $, where $K$ is the population size. According to the evolutionary algorithm, the child individuals in the population are iteratively generated and updated through several evolutionary operations (i.e., crossover, mutation and selection), which is elaborated in Algorithm \ref{alg:2}. Note that the fitness of each child architectures is directly evaluated through network inference. Once the termination criterion is met, the optimal backbone architectures are respectively obtained for different recognition tasks.

\begin{algorithm}[htb]
\caption{Evolutionary backbone search}
\label{alg:2}
\begin{algorithmic}[1]
\REQUIRE fine-tuned supernet, searching dataset ${D_{searching}}$, population size $K$, maximum number of generations $G$, crossover ratio $c$, mutation ratio $m$,
\STATE ${P_0} \leftarrow $ initialize a population with a size of $K$;

\FOR{$t = 1, ...,G$ }
\FOR{each individual in ${P_t}$}
\STATE Calculate the fitness of individual through the inherited weights from the supernet on ${D_{searching}}$;
\ENDFOR
\STATE Generate offspring $|{Q_t}| = K$ by crossover and mutation operations;
\FOR{each individual in $|{Q_t}|$}
\STATE Calculate the fitness of individual through the inherited weights from the supernet on ${D_{searching}}$;
\ENDFOR
\STATE ${R_t} \leftarrow {P_t} \cup {Q_t}$
\STATE ${P_{t + 1}} \leftarrow $ select $K$ individuals from ${R_t}$ by environment selection;
\ENDFOR
\ENSURE The searched backbone for the target task.
\end{algorithmic}
\end{algorithm}

\section{Experimental Setup}\label{section:4}
In this section, the experimental settings for evaluating the proposed RSBNet are presented. Section \ref{section:4.1}  gives a detailed description of the used RSI datasets. The construction of a new large scale RSI dataset is discussed in Section \ref{section:4.2} . For different recognition tasks, we give the evaluation metrics in Section \ref{section:4.3} . In Section \ref{section:4.4} , we present the implementation details of the proposed RSBNet.

\subsection{Data Sets Description}\label{section:4.1}
In our experiments, seven public RSI datasets are used for evaluation the performance of proposed methods in three RSI recognition tasks, including AID dataset \cite{xia2017aid}, NWPU-RESISC45 dataset \cite{cheng2017remote}, PatternNet dataset \cite{zhou2018patternnet}, RSI-CB dataset \cite{li2017rsi}, UC-Merced dataset \cite{yang2010bag}, ISPRS 2-D \cite{ISPRS} and NWPU VHR-10 dataset \cite{cheng2016survey}. Collected from various satellite sensors over diverse locations of the ground surface and under different conditions, these datasets contain rich scene categories, geographic structures and image variations. The first four datasets are built for RSI scene classification task; the NWPU VHR-10 dataset is a geospatial dataset used for multiclass object detection; the ISPRS 2-D dataset is constructed for the task of land-cover classification. A brief introduction to each dataset is described as below.
\begin{enumerate}[(1)]
  \item
  AID dataset: It consists of 10000 $600 \times 600$ aerial images of 30 scene categories, and the number of sample images in each class varies from 220 to 420.
  \item
  NWPU-RESISC45 dataset: This dataset contains totally 31500 sample images which are divided into 45 scene categories. Each scene category consists of 700 images with a fixed size of $256 \times 256$ pixels.
  \item
  PatternNet dataset: It contains 38 scene categories, and each category consists of 800 images of size $256 \times 256$ pixels.
  \item
  RSI-CB dataset: This dataset contains 35 scene categories of approximately 24,000 images. Each class contains an average of approximately 690 images with a size of $256 \times 256$ pixels.
  \item
  UC-Merced dataset: This small-scale dataset is a widely used scene classification benchmark dataset, which consists of 21 scene classes with 100 sample images in each class. These scene images are all with a size of $256 \times 256$.
  \item
  ISPRS 2-D dataset: This dataset is a sub-decimeter resolution dataset over the two cities of Vaihingen and Potsdam. The Vaihingen dataset contains 33 patches of different sizes (average $2494 \times 2064$), and the Potsdam dataset consists of 38 tiles ($6000\times 6000$ pixels). All pixels of the sample images are annotated with six common land cover categories.
  \item
  NWPU VHR-10 dataset: This dataset is a multi-source and multi-resolution object detection dataset covering 10 categories. The images are divided into two sub-datasets, including a positive sample dataset of 650 images and a negative sample dataset of 150 images.
\end{enumerate}

\subsection{Construction of a Large-Scale Dataset}\label{section:4.2}
\begin{figure}[htbp]
\centering
\includegraphics[scale=0.6]{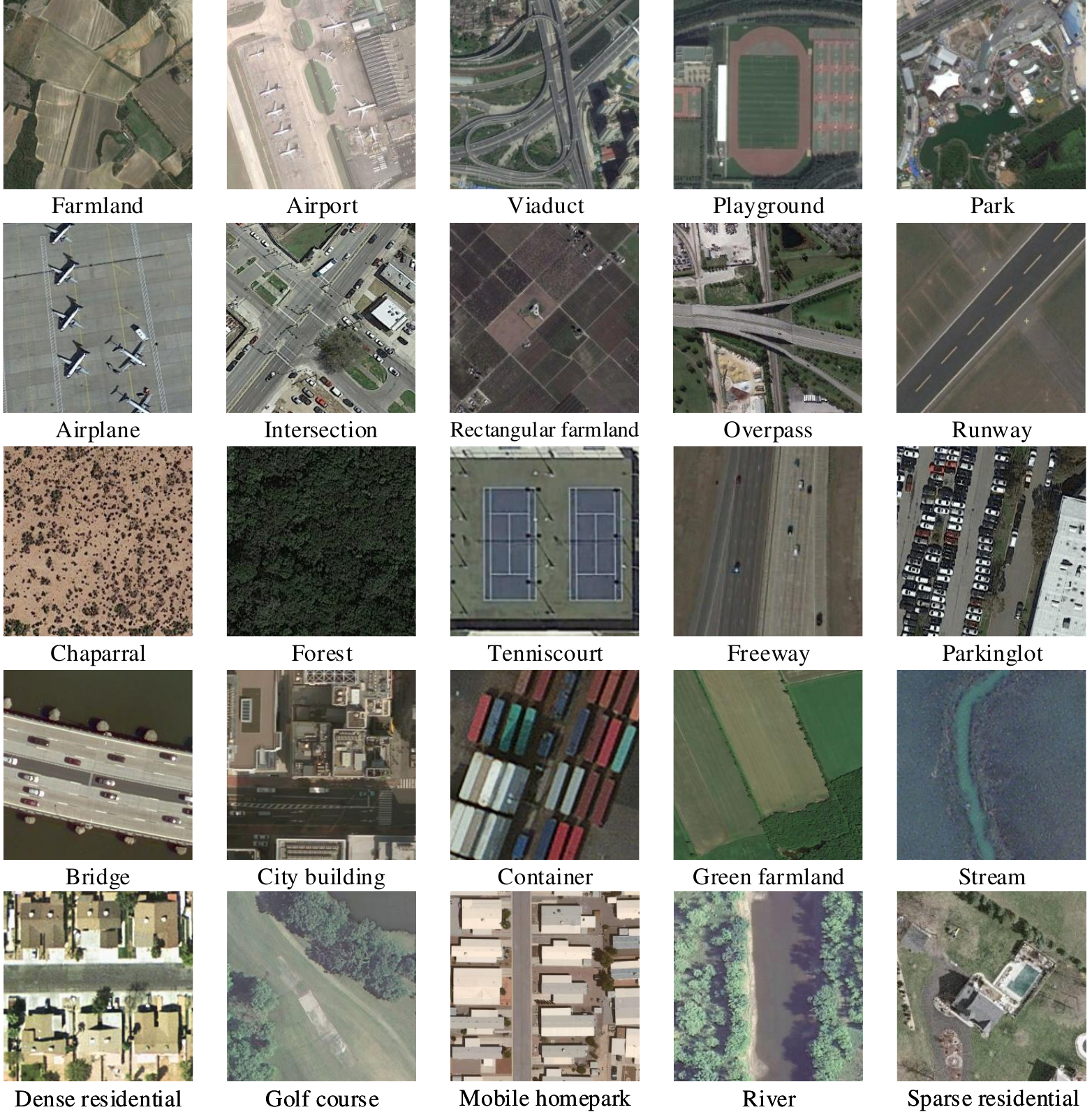}
  \caption{Illustration of some example images of the merged dataset VHRRS.}
  \label{fig:scene}
\end{figure}

The purpose of the supernet pretraining stage is to train a robust architecture that can be transferred to different recognition tasks. Considering the shortage and the significant specificity of RSI data, we define some mergence rules to build a new large-scale remote sensing image data set called VHRRS dataset. Since the remote sensing scene images usually contain a diversity of scene categories, land-cover categories and ground object categories, the heterogeneous data sources come from four large-scale remote sensing scene datasets: AID, NWPU-RESISC45, PatternNet, and RSI-CB. According to the idea of multi-duplication and multi-complementary, partial image data from these four datasets are selected to construct the VHRRS dataset, which contains 93068 high-resolution RSIs with 81 scene categories. Fig.\ref{fig:scene} shows some example images from VHRRS. In our experiments, the one-shot supernet is pretrained over the VHRRS dataset containing a diversity of heterogeneous domain information of geographical structures, making it feasible to find an excellent supernet with strong generalization ability for different recognition tasks.

\subsection{Evaluation Metrics}\label{section:4.3}
In our experiments, different quantitative evaluation metrics are adopted for different recognition tasks. In terms of the task of scene classification, two common evaluation metrics, namely overall accuracy (OA) and the mean Average Precision (mAP), are used to measure the final classification accuracy and quantify the performance of the searched backbone network.

For the land-cover classification task in ISPRS 2-D dataset, we used the evaluation criteria, including OA (overall accuracy), F1 score and pixel-wise accuracy of each categories. The OA is a global evaluation indicator, which calculates the proportion of correctly classified pixels. As a class-specific metric, F1 score is independent of the size of classes and calculated by the harmonic mean between precision and recall.

The accuracy of RSI object detection task is usually evaluated by precision and recall. Precision calculates the percentage of the predicted results that are correct. Recall involves the total number of positive instances that are correctly predicted, which reflects the model ability to correctly predict positive sample. Average precision (AP) is a common evaluation metric in object detection works, which are derived from precision and recall. Generally, AP is separately calculated for each class. As the average of the AP values of all object categories, the mean AP (mAP) is adopted as the final metric for evaluating overall accuracy.

\subsection{Implementation Details}\label{section:4.4}
As discussed in section \ref{section:3} , our proposed RSBNet search framework contains three basic stage: one-shot supernet pretraining, fine-tuning the supernet for target tasks on target dataset, and optimal backbone network search. In this section, we provide the implementation details of the three stages, all the experiments are implemented with Pytorch \cite{paszke2017automatic} framework using two NVIDIA GTX 2080Ti GPUs.

\begin{enumerate}[(1)]
  \item
  One-shot supernet pretraining: In the self-assembled dataset VHRRS into, 50\% of the images from AID and NWPURESISC45 and all of the images from PatternNet and RSI-CB are used for supernet pretraining. Common data augmentation operations are conducted on the training data to alleviate over-fitting problem, \emph{i.e.}, $224 \times 224$ patches are randomly cropped from the raw images with random mirroring and rotation. In the training stage, we train the supernet for 150 epochs using a batch size of 128. The initial learning rate is set to be 0.045 and the cosine learning rate decay is same as that in \cite{loshchilov2016sgdr}. The weights are updated by a SGD optimizer with a momentum of 0.9 and a weight decay of $4 \times {10^{ - 5}}$. The number of sampled subnetworks in each step is set to 5.

 \item
  Supernet fine-tuning: The pertained supernet is used as the backbone network and respectively transferred to three recognition tasks. The training settings are similar to those of the pretraining except that the training epoch is 50 and the initial learning rate is 0.01, and the channel number search is included. For the branch of scene classification, the pretrained supernet is fixed and directly fine-tuned on the target dataset. Since three datasets are used for performance evaluation, \emph{i.e.}, AID, NWPU-RESISC45 and UC-Merced, where the datasets AID and NWPU-RESISC45 are contained in the backbone dataset, the stage of fine-tuning for them is omitted. For the dataset UC-Merced, 50\% of the samples are used for fine-tuning. For the branch of land-cover classification, the supernet is used as the backbone network and combined with ASPP module \cite{chen2018encoder} to form the complete models for segmentation task. Following the works in \cite{chen2018encoder}, atrous convolutions are introduced in the backbone network and similar training settings are adopted. In terms of the ISPRS 2-D dataset, 11 images and 17 images are respectively selected from Vaihingen and Potsdam, and used as the training set to fine-tune the networks. We crop the large images into a series of small image patches of $224 \times 224$ pixels in order to adapt batch training. For the branch of object detection, the backbone network is separated from the fine-tuned supernet and equipped with detection head. For the dataset NWPU VHR-10, 60\% of samples are used for fine-tuning. The training settings are similar to \cite{shi2020global}.

  \item
  Backbone network search with EA: The third step is to conduct the backbone network search on the fine-tuned supernet through evolutionary controller. The search process is respectively conducted on the validation data from different datasets. For the datasets AID and NWPURESISC45, 20\% of the images (not used in pretraining) are used for search. For the dataset UC-Merced, 30\% of the samples are used for fine-tuning. For ISPRS 2-D dataset, 5 images and 7 images are respectively selected from Vaihingen and Potsdam, and used for search. For the dataset NWPU VHR-10, 20\% of samples are used for backbone search. In terms of evolutionary search, we maintain a population with 50 different architectures and the evolution process is repeated for 30 iterations. The crossover ratio is set to 0.5 and the mutation ratio is 0.25.
\end{enumerate}

Once the optimal backbone networks are obtained by evolutionary search, we retrain the searched backbone architectures with all the target training data based on the fine-tuning schedule, and then perform evaluation on the target test data. The training settings are similar those of the supernet fine-tuning except that the searched architecture is trained for 100 epochs.

\section{Results and Analysis}\label{section:5}
In this section, the experimental results of the searched backbones for different RSI recognition tasks are presented, including scene classification, land-cover classification and object detection. At first, we provide a discussion about the search cost, and then the visual results of the searched backbone networks are presented. Next, the recognition performances of the searched backbones are respectively evaluated on three tasks. Finally, we conduct ablation studies on the proposed approaches.

\subsection{Analysis of Search Cost}\label{section:5.1}
Since NAS algorithms usually consumes expensive computational resources, the search cost is a matter of concern in our proposed methods. In this section, we report the time cost of our method in different stages. Table \ref{tab:cost} reports the results for different datasets. Note that the time cost is measured based on the search space without channel number search. We conduct three individual runs for the pretraining stage and the time costs are listed in column 2. We can find that the time cost of different individual runs is similar, which indicates the pretraining process is stable due to the effectiveness of the proposed ensemble single-path training strategies. Since the datasets AID and NWPU-RESISC45 are contained in the backbone dataset, the fine-tuning stage of which is omitted. For the object detection task on NWPU VHR-10 dataset, we only report the time cost based on one-stage detection framework (\emph{i.e.}, RetinaNet). Compared with other two stages, the process of fine-tuning clearly uses less time since the network weights are properly initialized. In terms of the search stage, the tasks of land-cover classification and object detection take more time and computation cost than the task of scene classification. In practice, our methods are also more efficient and convenient to use because the training stage and the search stage are decoupled and relatively independent, so that the pretrained supernet can be transferred to different recognition tasks or different datasets.

\begin{table}[htbp]
\footnotesize
\centering
\renewcommand\arraystretch{1.5}
\renewcommand\tabcolsep{2.0pt}
\caption{The time cost of three stages on different datasets}
\begin{tabular}{c|c|c|c}
\hline
\multirow{2}{*}{Dataset}&Pretraining time& Fine-tuning time &Search time \\
 & (GPU days) & (GPU days) & (GPU days) \\
\hline
AID &  & 0 &0.8 \\
\cline{1-1}
\cline{3-4}
NWPU-RESISC45&Run 1: 1.8  & 0 & 1.0 \\
\cline{1-1}
\cline{3-4}
UC-Merced & Run 2: 1.7& 0.1 & 0.2 \\
\cline{1-1}
\cline{3-4}
Vaihingen& Run 3: 1.7 & 0.5 & 1.4 \\
\cline{1-1}
\cline{3-4}
Potsdam & & 0.6 & 1.6 \\
\cline{1-1}
\cline{3-4}
NWPU VHR-10 & & 0.3 &1.2 \\
\hline
\end{tabular}
\label{tab:cost}
\end{table}

\subsection{Search Results of Backbone architectures}\label{section:5.2}
\begin{figure}[htbp]
\centering
\includegraphics[scale=0.35]{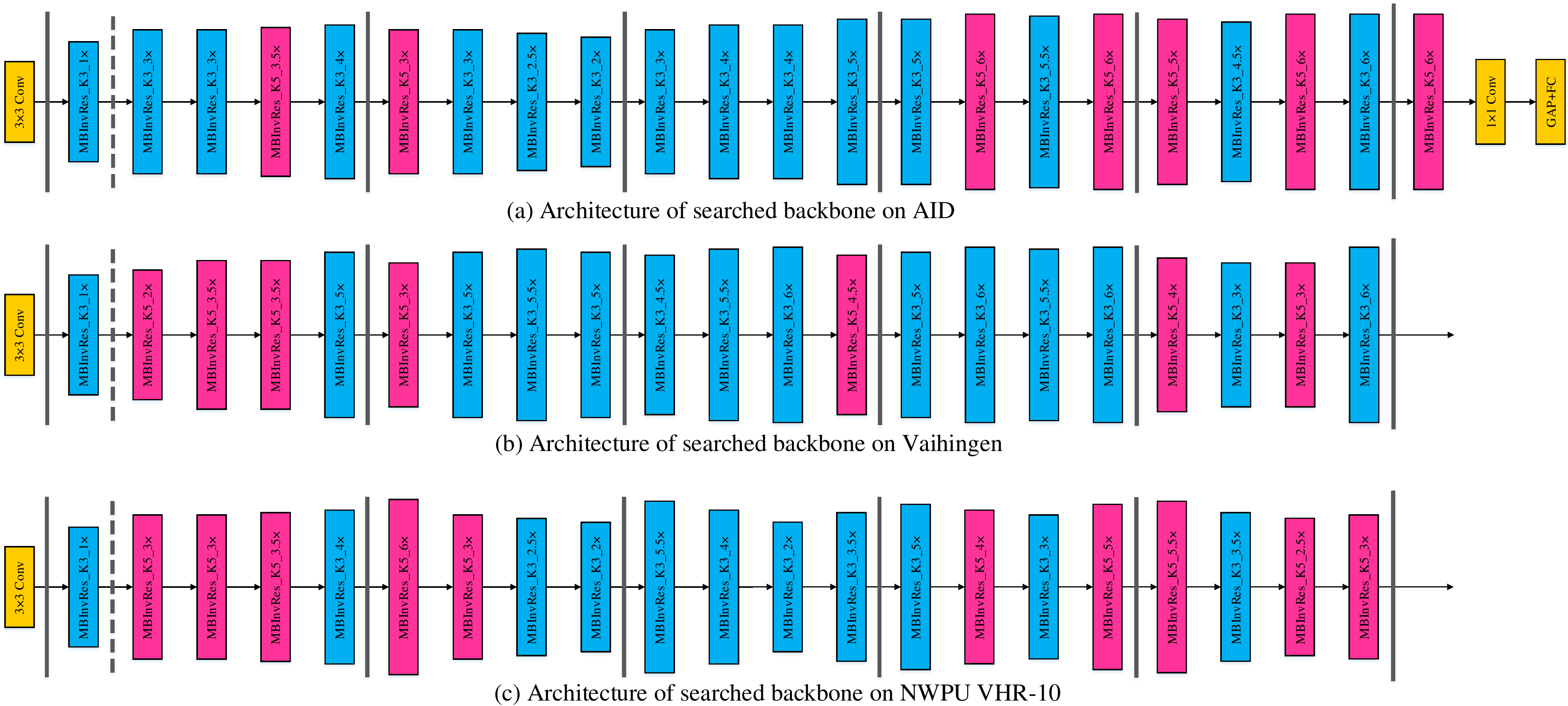}
  \caption{Illustration of the searched backbone architectures for different target datasets.}
  \label{fig:backbone}
\end{figure}

In this section, we report the architectures of the searched backbone networks on three datasets including AID (for scene classification), Vaihingen (for land-cover classification) and NWPU VHR-10 (for object detection). The detailed architectures are illustrated in Fig. \ref{fig:backbone}. The first two layers of the architecture have fixed operators and the intermediate layers are divided into multiple stages. There are six stages that can be searched in our search space and the basic structure of the candidate blocks is represented as ``MBInvRes''. The architecture in the top of Fig. \ref{fig:backbone} is RSBNet for AID dataset. The middle one is searched on Vaihingen dataset and the bottom one is searched with Faster R-CNN framework on NWPU VHR-10 dataset. ``MBInvRes\_Kx\_y$ \times$'' means a kernel size of x and an expansion rate of y for its depthwise convolution. We can find that the architecture searched for scene classification prefer to large expansion rates and large kernels at the high-level layers, which enables the full extractions of high-level features. However, the architecture searched for land-cover classification shows different patterns at the intermediate stages, where the small kernel and large expansion rate are preferred. We infer that mid-level layers contain more local semantic features. In terms of the backbone network for object detection, there are more large-kernel blocks in low-level layers, and the large expansion rate is usually used after a down-sampling operation. Based on these observations, we find that the backbone networks searched for different recognition tasks show significant differences and this specificity leads to different feature extraction abilities.

\subsection{Evaluation Results and Analysis}\label{section:5.3}
Through the switchable recognition module, the backbone networks are obtained for different recognition tasks. In this section, we report the evaluation results on three remote sensing scene datasets (\emph{i.e.}, AID, NWPU-RESISC45 and UC-Merced), a land-cover classification dataset (\emph{i.e.}, ISPRS 2-D) and an object detection dataset (\emph{i.e.}, NWPU VHR-10). Moreover, we can obtain the backbone with different model sizes through the channel number search. Therefore, we provide three versions of RSBNet under different computational constraints, \emph{i.e.}, RSBNet-small ($\leq$500M FLOPs), RSBNet-medium ($\leq$800M FLOPs) and RSBNet-large ($\leq$1G FLOPs).

\subsubsection{Scene classification results}\label{section:5.3.1}
In this set of experiments, we compare the performance of RSBNet with other three widely used manually designed CNN backbones, including VGG-16 \cite{simonyan2014very}, GoogLeNet \cite{szegedy2015going}, and ResNet-50 \cite{he2016deep}. We report the performance of different classical backbone in two cases: full training and fine-tuning on the pretrained model. The classification results are listed in Table \ref{tab:cls}, where the classification results of three classical backbones without ImageNet pretraining weights are listed in the second to fourth row. It can be found that both mAP and OA on datasets AID and NWPU-RESISC45 are lower than 90\%, which are poor and unacceptable in practice. However, the classification results of the pretrained models achieve significant improvements. The differences between these results also indicate the effectiveness of the pretrained weights. The classification accuracies obtained by the searched backbone are listed in the last three rows. We can see that the searched backbones obtain higher classification performance on both OA and mAP for these three datasets, especially for the NWPU-RESISC45 daset, RSBNet-large achieves accuracy improvement of 3$\sim$11\% compared with the classical manually designed CNN models. In addition, our RSBNet is able to search more compact architectures under different constraints. From the above results, we infer that the generalization capability of the pretrained models is limited in RSI field due to the complexity and specificity of RSIs. In contrast, the proposed RSBNet is proved to be effective for automatically designing suitable backbone network.

\begin{table}[t]
\centering
\footnotesize
\renewcommand\arraystretch{1.5}
\renewcommand\tabcolsep{2.0pt}
\caption{Scene classification results of the searched backbones and other backbones on three datasets}
\begin{tabular}{c|c|c|c|c|c|c|c}
\hline
\multirow{2}{*}{Backbone}&FLOPs& \multicolumn{2}{c|}{AID}& \multicolumn{2}{c|}{NWPU-RESISC45} &\multicolumn{2}{c}{UC-Merced}  \\
\cline{3-8}
&(G)& mAP(\%) & OA(\%) & mAP(\%) & OA(\%) & mAP(\%) & OA(\%) \\
\hline
full-trained VGG-16\cite{simonyan2014very} &15.55 &87.97&88.53&85.28&85.89&92.15&92.54\\
full-trained GoogLeNet\cite{szegedy2015going} & 2.01&84.84&84.95&83.07&83.29&91.59&91.68\\
full-trained ResNet-50\cite{he2016deep} &4.05 &91.57&92.05&89.97&91.32&93.43&93.42\\
\hline
VGG-16 (pretraining+fine-tuning)\cite{simonyan2014very} &15.55 &91.54&92.10&89.90&91.32&94.56&94.62\\
GoogLeNet (pretraining+fine-tuning)\cite{szegedy2015going} &2.01 &89.65&90.27&88.45&89.42&93.02&93.15\\
ResNet-50 (pretraining+fine-tuning)\cite{he2016deep} &4.05 &91.95&92.11&90.92&91.63&95.17&95.24\\
\hline
RSBNet-small &0.45&94.68&94.82&93.35&94.42&97.88&97.92\\
RSBNet-medium & 0.76&94.80&95.01&93.39&94.49&97.92&98.05\\
RSBNet-large &0.98&\textbf{94.85}&\textbf{95.06}&\textbf{93.42}&\textbf{94.52}&\textbf{98.12}&\textbf{98.20}\\
\hline
\end{tabular}
\label{tab:cls}
\end{table}

\subsubsection{Land-cover classification results}\label{section:5.3.2}
\begin{figure}[htb]
\centering
\includegraphics[scale=0.4]{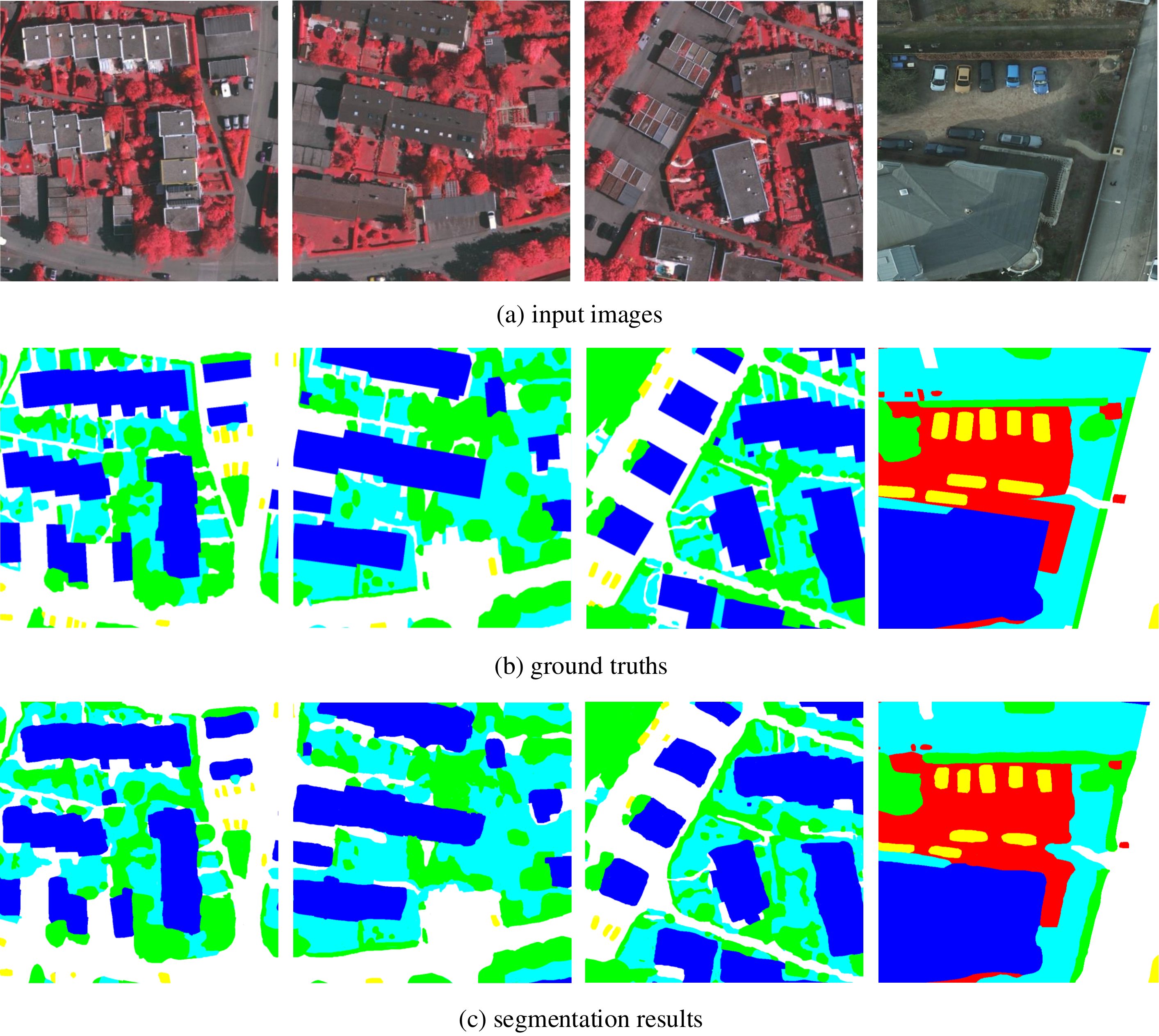}
  \caption{Example land-cover classification maps of ISPRS 2-D dataset. Annotation:White: impervious surfaces. Blue: buildings. Cyan: low vegetation. Green: trees. Yellow: cars. Red: clutter, background.}
  \label{fig:segmentation}
\end{figure}

In this set of experiments, we report the performance of the proposed methods on the task of remote sensing land-cover classification. We conduct the experiments on ISPRS 2-D dataset and compare the searched backbone to five competitive semantic segmentation models including FCN\cite{long2015fully}, UNet \cite{ronneberger2015u}, SegNet \cite{badrinarayanan2017segnet}, DeepLab v3 \cite{chen2017rethinking} and DeepLab v3+ \cite{chen2018encoder}. The classification results are shown in Table \ref{tab:seg1} and \ref{tab:seg2}. We can find that the RSBNet of different model size achieves satisfactory performance on different categories under different evaluation metrics. For the Vaihingen dataset, the category ``Building'' obtains the highest classification accuracy of 95.18\%. However, the category ``car'' gains the relatively poor classification accuracy of about 76\%, which indicates that the small objects with dense distribution are usually difficult to be correctly recognized. The fifth and sixth row of the Table \ref{tab:seg1} respectively reports the results of the models DeepLab v3 and DeepLab v3+, we can find that the accuracies of all categories are much better than that of first three models, especially the improvements on the indistinguishable categories ``car'', exceed about 5\% since the multi-scale context are captured by introducing the atrous convolution. As expected, the land-cover classification performance of our RSBNet on Potsdam dataset is better than that of other five models, which are based on handcrafted backbone network. For the visual results, we select four images from the validation set as an evaluation. The visual results of our method are illustrated in Fig.\ref{fig:segmentation}. It can be seen that our method provides an excellent preservation of the object boundaries and geometry. We also observe that the searched backbone can correctly classify both large multiscale objects (i.e., buildings) and small ground objects (\emph{i.e.}, cars) that are densely distributed.

\begin{table}[htb]
\centering
\footnotesize
\renewcommand\arraystretch{1.5}
\renewcommand\tabcolsep{2.0pt}
\caption{Land-cover classification results of the searched backbone and other methods on Vaihingen dataset}
\begin{tabular}{c|c|c|ccccc|c|c}
  \hline
  Model&Backbone&FLOPs(G)&Imp.surf& Building&Low veg. & Tree & Car & F1 & OA \\
  \hline
  FCN-8s\cite{long2015fully} & VGG-16 &15.55 &89.20 & 92.15 & 78.29 & 86.48 & 64.35 & 82.72 & 87.55 \\
  U-Net\cite{ronneberger2015u} & VGG-16 &15.55 &89.56 & 93.24 & 79.05 & 86.18 & 68.75 & 83.57 & 87.95 \\
  SegNet\cite{badrinarayanan2017segnet} & VGG-19&19.87 &90.14 & 92.37 & 77.67 & 86.54 & 68.99 & 83.75 & 87.42 \\
  DeepLabv3\cite{chen2017rethinking} &ResNet-50 &4.05 &91.22 & 92.89 & 80.57 & 88.05& 73.52&84.52 &88.62\\
  DeepLabv3+\cite{chen2018encoder} & ResNet-50&4.05 &92.20 & 93.42 & 80.86 & 88.16& 78.50&86.58 &88.79\\
  RSBNet-ASPP& RSBNet-small &0.48 &92.27 & 94.16 & 81.02 & 89.57 & 76.12 & 86.62 & 88.95 \\
  RSBNet-ASPP& RSBNet-medium &0.78 &92.35 & 95.02 & 81.18 & 89.66 & 76.15 & 86.69 & 89.02 \\
  RSBNet-ASPP& RSBNet-large &0.99 &\textbf{92.52} & \textbf{95.18} & \textbf{81.28} & \textbf{89.78} & \textbf{76.30} & \textbf{86.77} & \textbf{89.25} \\
  \hline
\end{tabular}
\label{tab:seg1}
\end{table}

\begin{table}[htb]
\centering
\footnotesize
\renewcommand\arraystretch{1.5}
\renewcommand\tabcolsep{2.0pt}
\caption{Land-cover classification results of the searched backbone and other methods on Potsdam dataset}
\begin{tabular}{c|c|c|cccccc|c|c}
  \hline
  Model&Backbone&FLOPs(G)& Imp.surf& Building&Low veg. & Tree & Car &Clutter& F1 & OA \\
  \hline
  FCN-8s\cite{long2015fully} & VGG-16 &15.55& 84.38 & 90.33 & 81.07 & 76.84 & 68.39 & 68.64 & 78.36 & 85.05 \\
  U-Net\cite{ronneberger2015u} & VGG-16 &15.55& 85.82 & 91.57 & 79.05 & 77.64 & 68.94 & 69.67 & 79.92 & 86.12\\
  SegNet\cite{badrinarayanan2017segnet} & VGG-19&19.87& 84.62 & 91.43 & 80.64 & 77.24 & 68.57 & 68.73 & 79.32 & 85.85 \\
  DeepLabv3+\cite{chen2018encoder} & ResNet-50&4.05&86.32 & 92.89 & 82.58 & 78.05& 75.58& 71.52 &83.56 & 87.39\\
  RSBNet-ASPP& RSBNet-small &0.49 & 89.89 & 94.32 & 84.96 & 85.94 & 76.96 & 73.84 & 85.31 & 88.79 \\
  RSBNet-ASPP& RSBNet-medium &0.79 & 89.92\ & 94.94 & 85.27 & 85.99 & 77.37 & 73.95 & 85.95 & 89.19 \\
  RSBNet-ASPP& RSBNet-large &0.98 &\textbf{90.91} &\textbf{95.33}& \textbf{85.64}& \textbf{86.29} & \textbf{78.45}& \textbf{74.42}& \textbf{86.61}& \textbf{89.89} \\
  \hline
\end{tabular}
\label{tab:seg2}
\end{table}

\subsubsection{Object detection results}\label{section:5.3.3}
\begin{figure}[htb]
\centering
\includegraphics[scale=0.7]{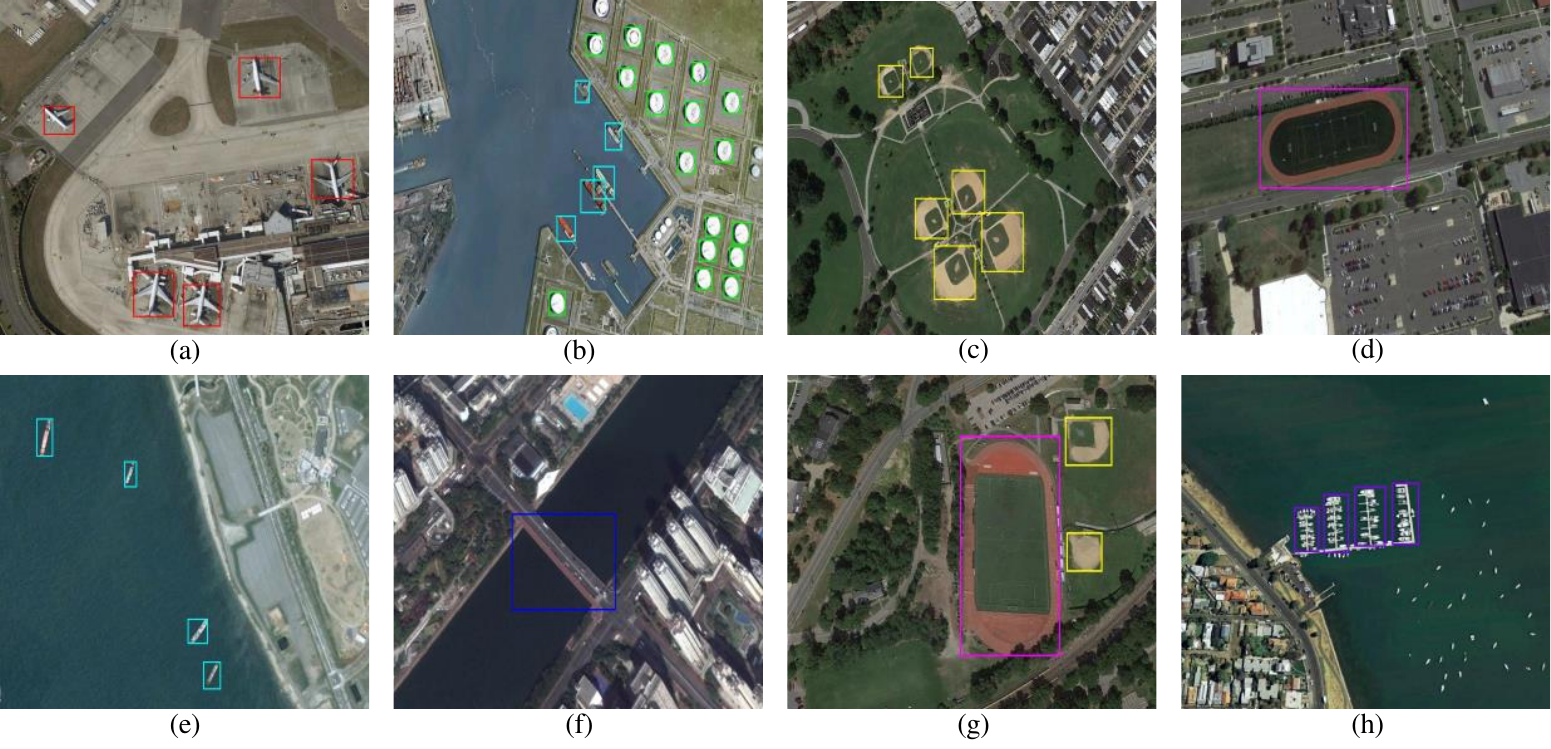}
  \caption{Example object detection visual results of NWPU VHR-10 dataset.}
  \label{fig:detection}
\end{figure}

In this section, we analyze the effects of the searched backbone on the performance of RSI object detection task. We evaluate the detection results of our RSBNet in different detectors (one-stage SSD and RetinaNet, two-stage Faster R-CNN and Mask R-CNN) on NWPU VHR-10 dataset. We do not constrain the model size of the searched backbones, and we embed the RSBNet into three detectors, \emph{i.e.}, RetinaNet (RSBNet-RN), Faster R-CNN (RSBNet-FR) and Mask R-CNN (RSBNet-MR).  The detection results of each category and mAP are shown in Table \ref{tab:det}. Through replacing the backbone in different detection frameworks by the searched backbone architecture and combining the FPN module, our methods obtain higher mAP than all compared models. Specifically, our embedded models RSBNet-RN, RSBNet-FR and RSBNet-MR achieve mAP improvement of 4.68\%, 1.69\% and 3.68\%, respectively, compared to their baseline models (\emph{i.e.}, RetinaNet with backbone ResNet50-FPN, Faster R-CNN with backbone ResNet50-FPN and Mask R-CNN with backbone ResNet50-FPN). In addition, our method also has a strong detection ability on the categories with a high proportion of small objects (\emph{i.e.}, ship and vehicle). In Fig.\ref{fig:detection}, we provide some object detection visual results on NWPU VHR-10 dataset to qualitatively evaluate the effectiveness of the searched backbone. It can be clearly observed that most ground objects are accurately and tightly covered by the predicted bounding boxes. Meanwhile, some ground objects of same classes with different sizes and texture information has also been correctly detected. Based on the quantitative analysis and qualitative results, we can conclude that the searched backbone network has an outstanding feature extraction and representation ability on the task of RSI object detection.

\begin{table}[htbp]
\centering
\scriptsize
\renewcommand\arraystretch{1.5}
\renewcommand\tabcolsep{1.0pt}
\caption{Object detection results of the searched backbone and other methods on NWPU VHR-10 dataset}
\begin{tabular}{c|c|cccccccccc|c}
  \hline

   \multirow{2}{*}{Model}&\multirow{2}{*}{Backbone}& Airplane& ship&Storage  &Baseball  & Tennis & Basketball  & Ground  &Harbor &Bridge &Vehicle &\multirow{2}{*}{mAP} \\
   & & & &tank &diamond  &court& court&track field & & & & \\
  \hline
  SSD\cite{liu2016ssd} &VGG-16               &77.45 &70.60 &76.20 & 89.30 &80.20 & 69.70 & 95.82 & 58.65 & 67.80 & 68.35 & 75.41 \\
  SSD-Lite\cite{sandler2018mobilenetv2} &MobileNetV2     &71.70 &68.60 &72.20 & 86.37 &78.29 & 68.62 & 94.15 & 52.63 & 59.76 & 65.25 & 71.76 \\
  RetinaNet\cite{lin2017focal} &ResNet50-FPN   &91.75 &86.69 &86.20 & 92.30 &85.26 & 79.70 & 96.87 & 79.98 & 87.80 & 84.35 & 87.09 \\
  RetinaNet\cite{lin2017focal} &ResNet101-FPN  &95.69 &88.63 &88.12 & 93.19 &88.29 & 85.63 & 98.82 & 82.65 & 89.95 & 86.29 & 89.73 \\
  Faster R-CNN\cite{ren2016faster} &VGG-16      &94.20 &85.62 &69.65 & 87.28 &89.14 & 79.77 & 79.82 & 68.65 & 77.86 & 82.45 & 81.45 \\
  Faster R-CNN\cite{ren2016faster} &ResNet50    &98.72 &89.85 &87.84 & 91.25 &91.53 & 89.92 & 92.68 & 85.30 & 82.92 & 84.07 & 89.41 \\
  Faster R-CNN\cite{ren2016faster} &ResNet50-FPN&99.12 &89.89 &88.98 & 92.13 &91.97	& \textbf{90.62}	& 93.11	& 86.52	& 82.51 & 84.49	& 89.93 \\
  Mask R-CNN\cite{he2017mask} &ResNet50-FPN  &99.14 &88.25 &85.94 & 92.43 &89.71	& 88.60	& 90.41	& 82.59	& 86.24 & 89.42	& 89.27 \\
  \hline
  RSBNet-RN&RSBNet-FPN      &99.29 &89.48 &91.16 & \textbf{93.25} &88.78	& 87.45	& \textbf{98.93}	& 91.41	& 89.78 & 88.17	& 91.77 \\
  RSBNet-FR&RSBNet-FPN      &99.08 &87.48 &95.10 & 91.69 &\textbf{92.28}	& 89.89	& 95.75	&\textbf{ 92.53	}& 85.78	& 86.63	& 91.62 \\
  RSBNet-MR&RSBNet-FPN      &\textbf{99.34} &\textbf{90.22} &\textbf{98.12} & 91.25 &86.78	& 89.45	& 96.82	& 92.41	& \textbf{95.42}	& \textbf{89.67}	& \textbf{92.95 }\\
  \hline
\end{tabular}
\label{tab:det}
\end{table}

\subsection{Ablation Studies}\label{section:5.4}
\subsubsection{Effectiveness of the supernet training strategy}\label{section:5.4.1}
\begin{figure}[htb]
\centering
\includegraphics[scale=0.25]{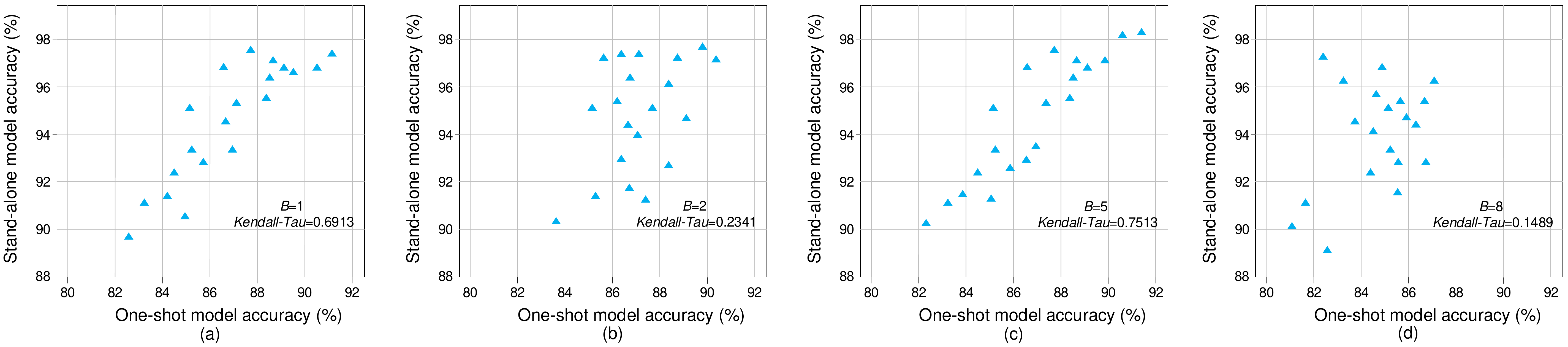}
  \caption{Comparison of one-shot and stand-alone model accuracies under different sampling strategies.}
  \label{fig:correlation}
\end{figure}

In this section, we conduct ablation experiments to validate the efficiency and the effectiveness of our ensemble single-path supernet training strategy introduced in section \ref{section:3.2} , and to investigate the sensitivity of the number of sampled paths. All the comparison experiments in this section are conducted on UC-Merced dataset and follow the pipeline and experimental settings described in section \ref{section:4.4} . The number of sampled single-path $B$ is an important hyperparameter that needs to be carefully tuned in our search framework. We set $B$ to 1, 2, 5 and 8, respectively, and conduct supernet pretraining process with different $B$. After the stage of supernet pretraining, we randomly sampled around 20 sing-path architectures from the supernet trained with different strategies, respectively. For these sampled architectures, we compute the classification accuracy of each one through inheriting the weights of the one-shot model. Then, we retrained each of the sampled architectures from scratch for 50 epochs with a batch size of 128. We visualize the accuracies of sampled architectures under different evaluation schemes in Fig.\ref{fig:correlation}. Furthermore, we adopt the \emph{Kendall Tau}\cite{kendall1938new} metric to measure the ranking correlation between the one-shot model accuracies and the stand-alone model accuracies. When $B=8$, we can see that the value of \emph{Kendall Tau} is only about 0.15, which means that the performance of stand-alone model trained from scratch has low ranking correlation with that of subnetworks directly extracted from supernet. RSBNet obtains the \emph{Kendall Tau} of 0.75 when $B=5$, which is better than other cases and shows the inherited weights are predictive for the accuracy. We infer that the mini batch single-path training stabilize the training procedure and make the pretrained supernet more robust. Therefore, we choose a reasonable $B=5$ as the best trade-off in our experiments.

\subsubsection{Effectiveness of the channel number search}\label{section:5.4.2}
As described in section \ref{section:3.3} , we train a slimmable supernet on target task in the stage of fine-tuning, which allows the channel number (\emph{i.e.}, the expansion ratio of the depthwise separable convolution)to be searched in the third stage, leading to better accuracy under constrained resources. In this section, we conduct two sets of comparison experiments on three datasets, including datasets AID, NWPU-RESISC45 and Vaihingen, to investigate the impact of channel number search on the performance of backbone. For the first set of experiments, the channel number configuration is excluded from the search space, and the expansion ratio of the depthwise convolution is fixed to full width (\emph{i.e.}, $e=6$). In the second set of experiments, the number of channels can be selected from the predefined width list (\emph{i.e.}, $e \in [2,6]$ (with stride 0.5))during the search process. To make a fair comparison, we set the computational constraint FLOPs$\leq$500M, and the number of search iterations for these two sets of experiments is set to be the same as the settings in section \ref{section:4.4} . Table \ref{tab:spa} shows the comparison results of different search space on three datasets. It can be seen that the searched backbones based on the joint space achieve better classification performance for different datasets, which indicates that proper layer width significantly improves the representation ability. In addition, the searched backbone can meet the target constraints more tightly through introducing channel number search, which means the final model can fully use the available computation resource.
\begin{table}[htb]
\centering
\scriptsize
\renewcommand\arraystretch{1.5}
\renewcommand\tabcolsep{2pt}
\caption{Comparison results of different search space on three datasets.}
\begin{tabular}{c|c|c|c|c|c|c|c|c|c}
\hline
& \multicolumn{3}{c|}{AID}& \multicolumn{3}{c|}{NWPU-RESISC45} &\multicolumn{3}{c}{Vaihingen}  \\
\cline{2-10}
& OA & Params&FLOPs & OA & Params &FLOPs & OA & Params &FLOPs\\
\hline
Search without channel number&94.28\%	&5.6M &476M	&94.06\%	&5.7M &485M	&88.65\%	&5.4M &489M\\

\hline
Search with channel number  &95.01\%	&5.9M &495M	&94.49\%	&5.5M &492M	&89.02\%	&5.9M &496M\\

\hline
\end{tabular}
\label{tab:spa}
\end{table}

\subsubsection{Necessity of the fine-tuning stage}\label{section:5.4.3}
\begin{figure}[htb]
\centering
\includegraphics[scale=0.4]{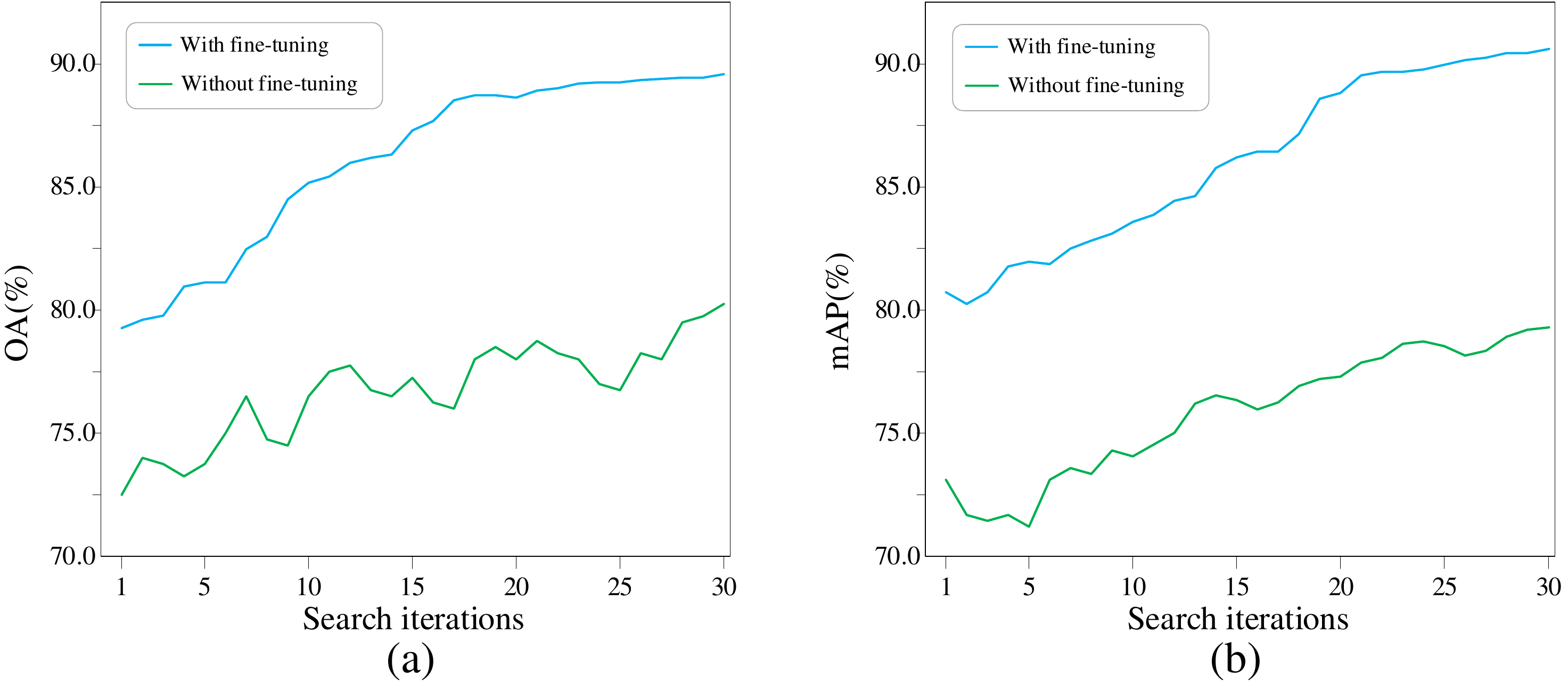}
  \caption{Change of best fitness value in the population during the search process. (a) Curve of OA in Vaihingen experiments; (b) Curve of mAP in NWPU VHR-10 experiments.}
  \label{fig:finetune}
\end{figure}

The fine-tuning stage aims to adjust the weights of the backbone network equipped with different recognition heads. Actually, some NAS works tend to search the architectures based on a proxy task and then transfer to other tasks or datasets. We demonstrate this manner is sub-optimal in our work through the ablation studies. Specifically, we conduct two sets of comparison experiments on two datasets, including datasets Vaihingen and NWPU VHR-10. In the first set of experiments, the pretrained backbone network without fine-tuning is directly transferred to the search stage of the two branches for land-cover classification and object detection respectively. In contrast, the second set of experiments is conducted with complete steps as described in section \ref{section:3} . Table \ref{tab:fine} shows the comparison results of these two sets of experiments. We can find that the recognition results show a relatively large gap between these two training schedules. The supernet fine-tuned based on target dataset achieves satisfactory performance, and which is much better than that of the supernet only trained based on VHRRS. Fig.\ref{fig:finetune} shows the change curve of best fitness value in the population during the search process over these two sets of experiments. The search process in the first set of experiment is unstable, and the best individual in the population is much poor than that in the second set of experiment, where the supernet is finetuned on the target training dataset. We infer that the best supernet searched on scene classification cannot provide proper initial model weights for other recognition tasks such as land-cover classification and object detection due the issue of domain shift.

\begin{table}[t]
\centering
\scriptsize
\renewcommand\arraystretch{1.4}
\renewcommand\tabcolsep{10pt}
\caption{Comparison results of different fine-tuning strategy on two datasets.}
\begin{tabular}{c|c|c}
\hline
& Vaihingen& NWPU VHR-10  \\

& OA(\%) & mAP(\%) \\
\hline
Without fine-tuning&80.51 & 79.19\\

\hline
With fine-tuning  &89.34 & 91.24\\

\hline
\end{tabular}
\label{tab:fine}
\end{table}

\begin{figure}[htbp]
\centering
\includegraphics[scale=0.4]{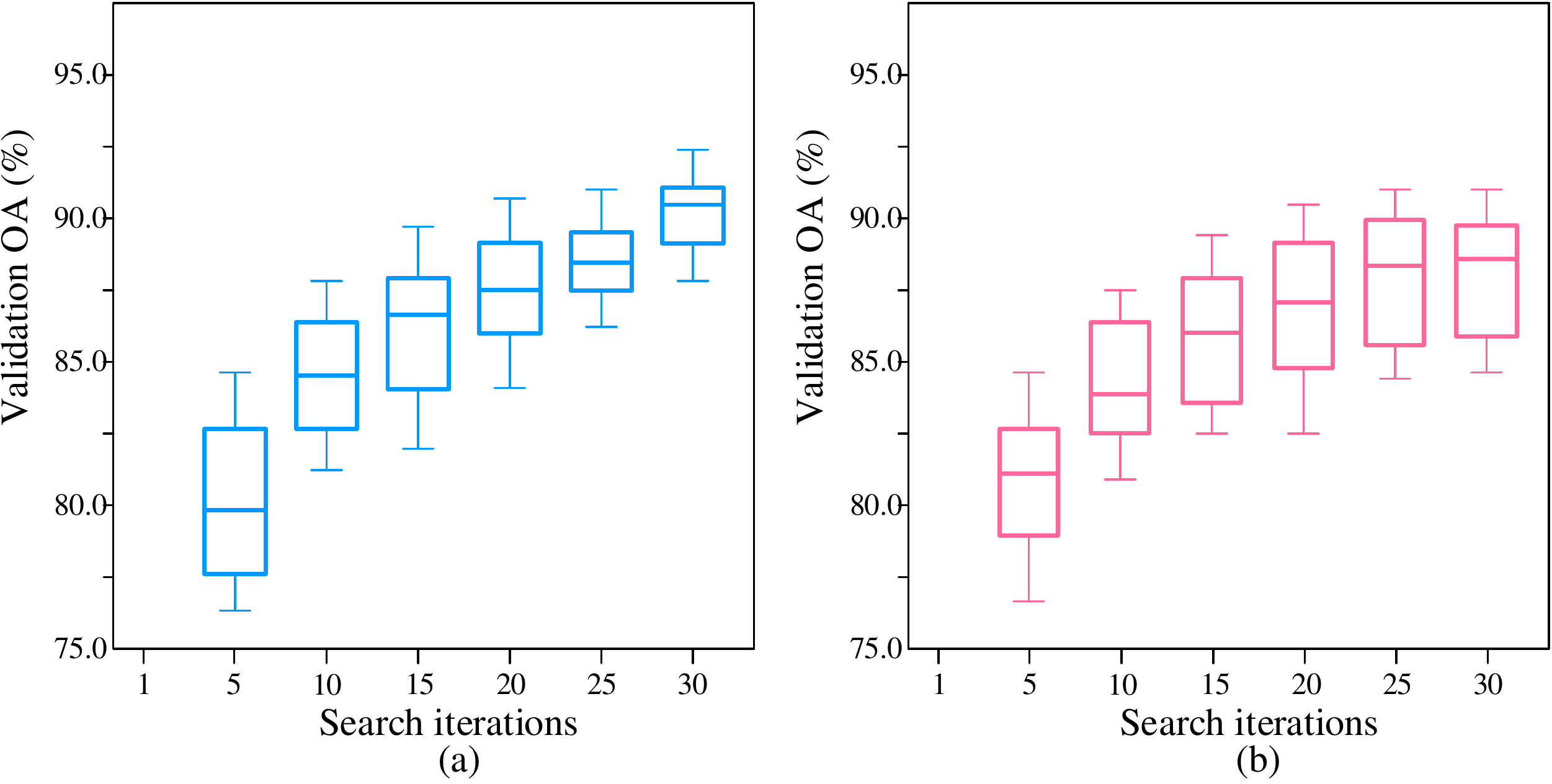}
  \caption{Change of validation accuracy during the search process. (a) Curve of evolutionary search; (b) Curve of random search.}
  \label{fig:random}
\end{figure}

\subsubsection{Effectiveness of the evolutionary search}\label{section:5.4.4}
In our approaches, any adequate search algorithm is practicable in the search stage since the supernet training stage and the architecture search stage are decoupled. To demonstrate the effectiveness of evolutionary search algorithm, we compare the performance difference of the searched architectures under different search strategies including evolutionary algorithm and random search. The comparison experiments are conducted on two datasets UC-Merced and Vaihingen. For the evolutionary search, the optimal architecture is obtained by the training settings described in section \ref{section:4.4} . In the case of random search, we random sample 50 different architectures from the fine-tuned supernet in each iteration. The weight sharing strategy are used in both cases, but the evolutionary search exploits the information between individuals through evolutionary operation. Fig.\ref{fig:random} plots the accuracy on the UC-Merced validation dataset during search, using evolutionary and random search methods, respectively. The overall accuracy results of two datasets are shown in Table \ref{tab:str}. It can be seen that the evolutionary searched architectures perform better than the random searched architectures, which proves the effectiveness of evolutionary search and also indicates that the random search is an alternative method to select competitive candidates from a large search space.

\begin{table}[htbp]
\centering
\scriptsize
\renewcommand\arraystretch{1.4}
\renewcommand\tabcolsep{12pt}
\caption{Comparison results of different search strategy on two datasets.}
\begin{tabular}{c|c|c}
\hline
& UC-Merced&Vaihingen  \\
&OA(\%)&OA(\%)\\
\hline
Random search&98.05 & 89.08\\

\hline
Evolutionary search &98.23 & 89.34\\

\hline
\end{tabular}
\label{tab:str}
\end{table}

\section{Conclusion}\label{section:6}
In this paper, we investigate a new design paradigm for the backbone architecture in remote sensing image recognition. A novel one-shot architecture search framework based on weight-sharing strategy is proposed, called RSBNet, which consists of three stages: one-shot supernet pre-training, supernet fine-tuning and candidate backbone search. The searched backbones are able to flexibly adapt to different recognition tasks including scene classification, land-cover classification and object detection. Various comparison experiments and ablation studies demonstrate the effectiveness of the proposed approaches. Our works raise up a new idea for various RSI recognition tasks. We also believe that the potential of NAS methods in the field of remote sensing will receive more attention.

\section*{Declaration of Competing Interest}
The authors declare that they have no known competing financial interests or personal relationships that could have appeared to influence the work reported in this paper.

\section*{Acknowledgements}
This research was funded by the National Natural Science Foundation of China under Grant 61772399, Grant U1701267, Grant 61773304, Grant 61672405 and Grant 61772400, the Key Research and Development Program in Shaanxi Province of China under Grant 2019ZDLGY09-05, the Program for Cheung Kong Scholars and Innovative Research Team in University Grant IRT\_15R53, and the Technology Foundation for Selected Overseas Chinese Scholar in Shaanxi under Grant 2017021 and Grant 2018021, and in part by the National Key Research and Development Program of China under Grant 2017YFC08219.

\end{spacing}

\bibliography{citation}
\end{document}